\documentclass[twocolumn]{aastex61}
\pdfoutput=1

\usepackage{morefloats}
\usepackage{epstopdf}
\usepackage{xcolor, graphicx}
\usepackage{amssymb}
\usepackage{amsmath}
\usepackage{longtable}
\usepackage{natbib}
\begin{document}


\title{A Likely Super Massive Black Hole Revealed by its Einstein Radius in Hubble Frontier Fields Images}

\author{Mandy C. Chen}\affiliation{Department of Physics, The University of Hong Kong, Pokfulam Road, Hong Kong}
\author{Tom Broadhurst}\affiliation{Department of Theoretical Physics, University of the Basque Country UPV/EHU, Bilbao, Spain}\affiliation{IKERBASQUE, Basque Foundation for Science, Bilbao, Spain}\affiliation{Donostia International Physics Center (DIPC), 20018 Donostia-San Sebastian (Gipuzkoa) Spain. }
\author{Jeremy Lim}\affiliation{Department of Physics, The University of Hong Kong, Pokfulam Road, Hong Kong}
\author{Jose M.~Diego}\affiliation{IFCA, Instituto de F\'isica de Cantabria (UC-CSIC), Av.~de Los Castros s/n, 39005 Santander, Spain}
\author{Youichi Ohyama}\affiliation{Institute of Astronomy and Astrophysics, Academia Sinica, Taipei 10617, Taiwan}
\author{Holland Ford}\affiliation{Department of Physics and Astronomy, Johns Hopkins University, Baltimore, MD 21218, USA}
\author{Narciso Ben\'itez}\affiliation{Instituto de Astrof\'isica de Andaluc\'ia (IAA-CSIC), Glorieta de la Astronom\'ia, 18008 Granada, Spain}

\correspondingauthor{Mandy C. Chen}
\email{chencc@connect.hku.hk}


\begin{abstract}

At cosmological distances, gravitational lensing can in principle provide direct mass measurements of supermassive black holes (SMBH).  Here, we directly estimate the mass of a SMBH in the brightest cluster galaxy (BCG) of MACS J1149.5+2223 at $z=0.54$ using one of the multiply-lensed images of a background spiral galaxy at $z=1.49$ projected close to the BCG.  A lensed arc is curved towards the BCG centre, corresponding to an intrinsically compact region in one of the spiral arms.  This arc has a radius of curvature of only $\sim0\farcs6$, betraying the presence of a local compact deflector.  Its curvature is most simply reproduced by a point-like object with a mass of $8.4^{+4.3}_{-1.8}\times10^{9}M_\odot$, similar to SMBH masses in local elliptical galaxies having comparable luminosities.  The SMBH is noticeably offset by $4.4\pm0.3$ kpc from the BCG light centre, plausibly the result of a kick imparted $\sim2.0\times10^7$ years ago during the merger of two SMBHs, placing it just beyond the stellar core.  A similar curvature can be produced by replacing the offset SMBH with a compact galaxy having a mass of $\sim2\times 10^{10}M_\odot$ within a cutoff radius of $<4$ kpc, and an unusually large $M/L>50(M/L)_\odot$ to make it undetectable in the deep Hubble Frontiers Fields image, at or close to the cluster redshift; such a lensing galaxy, however, perturbs the adjacent lensed images in an undesirable way.

\end{abstract}



\section{Introduction}
While the ubiquitousness of supermassive black holes (SMBHs) at the centres of relatively massive galaxies is widely accepted, the origin and growth of these enigmatic objects remain poorly understood if at all known. The now familiar $M_{BH}-\sigma$ relation suggests a co-evolution in mass between the SMBH and its host galaxy (e.g.~\citealt{Kormendy_Ho2013}), providing support for hierarchical mergers in structure formation theories. At very early times ($z>6.0$), however, surprisingly large SMBH masses of $\sim 10^9M_{\odot}$ have been inferred associated with host galaxies having relatively low masses (e.g.~\citealt{Wu2015,BennyScience}). Similarly, in the local Universe, SMBHs that lie well above the established $M_{BH}-\sigma$ relationship have been found, comprising perhaps ``fossil" cases with little past merging \citep{NGC1277_first,NGC1277_second}. As the accuracy of local black hole and host galaxy masses improves, the traditional $M-\sigma$ relations \citep{Ferrarese2000} have become more complex \citep{Graham2016review}. To understand the growth of SMBH masses over cosmic time, what is clearly needed is the ability to accurately measure their masses over a broad range of epochs.

To date, three widely used methods have been employed to measure SMBHs: modelling of stellar or gas kinematics, reverberation mapping, and scaling relations developed in large part from the results of reverberation mapping. For local galaxies where high spatial resolutions are possible, SMBH masses are determined through measurements of stellar or gas kinematics within a region where the gravitational force of the SMBH is dominant (e.g.~\citealt{Ferrarese&Ford1999,Kormendy2004,Meyer2012}). For more distant galaxies that have optically bright active galactic nuclei (AGNs), reverberation mapping can be used to infer the masses of their SMBHs. This approach requires measuring two parameters during a change (increase) in the brightness of an AGN. One of these parameters is the widths of emission lines from the broad-line region (BLR) of the AGN. The width of a given emission line is attributed to the orbital motion of that line-emitting gas surrounding the accretion disk of the SMBH. The second parameter is the time delay between a change in the continuum (from the accretion disk) and the emission-line (from the BLR) fluxes. This time delay corresponds to the light travel time between the accretion disk (assumed to be very small) and the region in the BLR at which a given emission line arises, and therefore the radius of the line-emitting region from the SMBH. From these two parameters, the mass of the SMBH can be derived. The major source of uncertainty in this method is the uncertainty in the BLR geometry. Because derivations of SMBH masses from reverberation mapping requires measurements with high signal-to-noise, this method has so far been restricted primarily to AGNs in nearby galaxies ($z<0.3$) \citep{Kaspi2005,Kaspi2007}. Results from reverberation mapping have been used to develop empirical scaling relations between the radius of BLR and the AGN optical luminosities based on different emission lines (e.g. H$\beta$ and Mg II), whereby the radius of the BLR have been found to scale with the AGN luminosity. This relationship implies a long time delay between changes in the continuum and emission-line fluxes at high AGN luminosities, making reverberation mapping impractical for these objects. Instead, the scaling relations developed from reverberation mapping have been used to infer SMBH masses in luminous AGNs and quasi-stellar objects (QSOs) at intermediate or high redshifts (see review by \citealt{Bentz2009}).

Gravitational lensing provides a promising new approach to directly measure SMBH masses -- one that does not depend on whether the SMBH is active or not, and furthermore that is almost irrespective of distance. The Einstein radius, $\theta_e$, of a point mass depends simply on the distances involved and scales slowly with mass: $\theta_e=0\farcs3(\frac{M}{10^{10}M_\odot})^{\frac{1}{2}}(\frac{D}{Gpc})^{-\frac{1}{2}}$ for a typical lens redshift of $z\simeq 0.5$ ($D=\frac{D_{l}D_{ls}}{D_{s}}$ where $D_{l}$, $D_s$ and $D_{ls}$ are the angular-diameter distances to the lens, the source, and between the lens and the source, respectively). At a limiting angular resolution of $\simeq 0\farcs1$, this Einstein radius is resolvable for a wide range in point masses of $M>10^8M_\odot$. In the situation where the background source and the foreground lensing object are closely aligned in the sky, a central de-magnified image is generic to this lensing geometry, such that the larger the SMBH mass the more this central image is attracted towards the SMBH and de-magnified \citep{mao2001,Rusin2005,Hezaveh2015}. 

One caveat of this approach, however, arises from the fact that when producing a certain magnification factor on the central lensed image, the SMBH mass is degenerate with the slope of the lensing galaxy's central mass profile, as is clearly illustrated in \cite{Hezaveh2015}. As a consequence, the presence (or absence) of a central image can only provide upper (or lower) limit of the SMBH mass. The first example was the multiply-lensed quasar PMN J1632-0033. Observations with the Very Long Baseline Array (VLBA) and Very Large Array (VLA) revealed a central image, thus constraining the mass of the central SMBH in the foreground lensing galaxy to be $M_{BH}<2\times10^8M_\odot$ \citep{Winn2003,Winn2004}. More recently, radio observations with the Atacama Large Millimeter/submillimeter Array (ALMA) of the lensed system SDP81 fail to detect a central image within the Einstein ring of the lensed background galaxy, thus placing a lower limit of $M_{BH}\sim3\times10^8M_\odot$ on the mass of a central SMBH in the foreground lensing elliptical galaxy at $z=0.3$ \citep{Tamura2015,Wong2015}. The SMBH masses thus derived for the foreground lensing galaxies in both PMN J1632-0033 and SDP81 are in agreement with the local $M_{BH}-\sigma$ relation. 

\cite{Quinn2016} discussed the implications for the non-detection of a central image in the lens system CLASS B1030+074 with the data from VLA and the extended Multi-Element Remote-Linked Interferometer (e-MERLIN), and argued in favour of a central SMBH with a mass slightly greater than that implied by the local $M_{BH}-\sigma$ relation, yet again, whether the SMBH is required in the lens model is dependent on the mass profile chosen for the lensing galaxy. A relatively large SMBH mass of $\sim1.2\times10^{10}M_\odot$, lying well above the $M_{BH}-\sigma$ relation, has been inferred for the brightest cluster galaxy (BCG; central giant elliptical galaxy) in the cluster Abell 1201 at $z=0.17$ based on the detection of a faint central image based on observations with the Hubble Space Telescope (HST) \citep{Smith2017}. In this case, however, the authors find that the observed parameters of the central image can be equally well explained by a cuspy stellar M/L ratio for the BCG. 

In this paper, we report a direct measurement of the mass of a SMBH through gravitational lensing. This SMBH is hosted by the BCG in the galaxy cluster MACSJ1149.5+2223 (hereafter MACS 1149) at a redshift of z = 0.543 \citep{Ebeling2007}. Unlike in the previous examples mentioned above, in this case the background lensed galaxy and BCG are not closely aligned in the sky. Instead, one of the multiply-lensed images of this background galaxy happens to be projected close to the BCG, which locally perturbs this image. One of the numerous compact HII regions in the background spiral galaxy is lensed into a curved arc, pointing to and betraying the presence of a SMBH. MACS 1149 is one of the six clusters from the Hubble Frontier Fields (HFF) program (PI: J. Lotz). Gravitational lensing by this cluster has been intensively studied (e.g.~\citealt{Zitrin&Broadhurst2009_MACS1149,Smith2009,Zheng2012.Nature,Rau2014,Sharon2015,Grillo2016}). The first (and so far only) multiply-lensed supernova was detected in this cluster \citep{Kelly2015,Oguri2015,Kelly2016,2016ApJ...817...60T}, and the discovery of a transient in a lensed image in this cluster has been attributed to microlensing by intracluster stars of a single star in a background lensed galaxy \citep{Kelly2018,Diego2018}. 

Understanding and reproducing the curved lensed arc in the image of the aforementioned background spiral galaxy involves the following separate steps: (1) deducing a robust cluster lens model for MACS 1149 and refining the mass model for the BCG so as to produce all the multiply-lensed images found towards the cluster, in particular those projected close to the BCG, as described in sections \ref{data} and \ref{lens_modelling}; (2) inferring the need for a local deflector to reproduce the curvature of L1, as described in section \ref{L1_investigation_section}; (3) determining the nature of this local deflector and its physical parameters, as described in sections \ref{result_section} and \ref{lenstool_section}. Our results are discussed in section \ref{discussion}, and a concluding summary is presented in section \ref{conclusion}.


\section{data}\label{data}
\subsection{HFF and CLASH}
We retrieved the already reduced HFF public imaging data (epoch2) for MACS 1149 taken with the Hubble ACS and WFC3 (PI: J. Lotz) from the Mikulski Archive for Space Telescopes (MAST)\footnote{\url{https://archive.stsci.edu/pub/hlsp/frontier/macs1149/images/hst/v1.0-epoch2/}}. From the HFF, there are a total of 140 orbits in the optical and infrared bands (filters: F435W, F606W, F814W, F105W, F125W, F140W and F160W) devoted to MACS 1149. We used this data for constructing the cluster lens model. MACS 1149 also is one of the clusters observed in the Cluster Lensing And Supernovae Survey with Hubble (CLASH) program (PI: M. Postman), for a total of 18 orbits in the UV, optical, and infrared bands spanning 17 filters\footnote{\url{https://archive.stsci.edu/prepds/clash/}}. For the purpose of this study, the F435W data from CLASH (1.5 orbits) was also added to the F435W data from HFF (18 orbits), albeit leading to only a small improvement of the signal-to-noise ratio due to the significantly shorter exposure time of CLASH. 

\begin{figure*}[tp!]
\centering
\graphicspath{{/Users/Mandy/GoogleDrive/BH_paper/ApJ/figures/}}
\includegraphics[width=13.5cm]{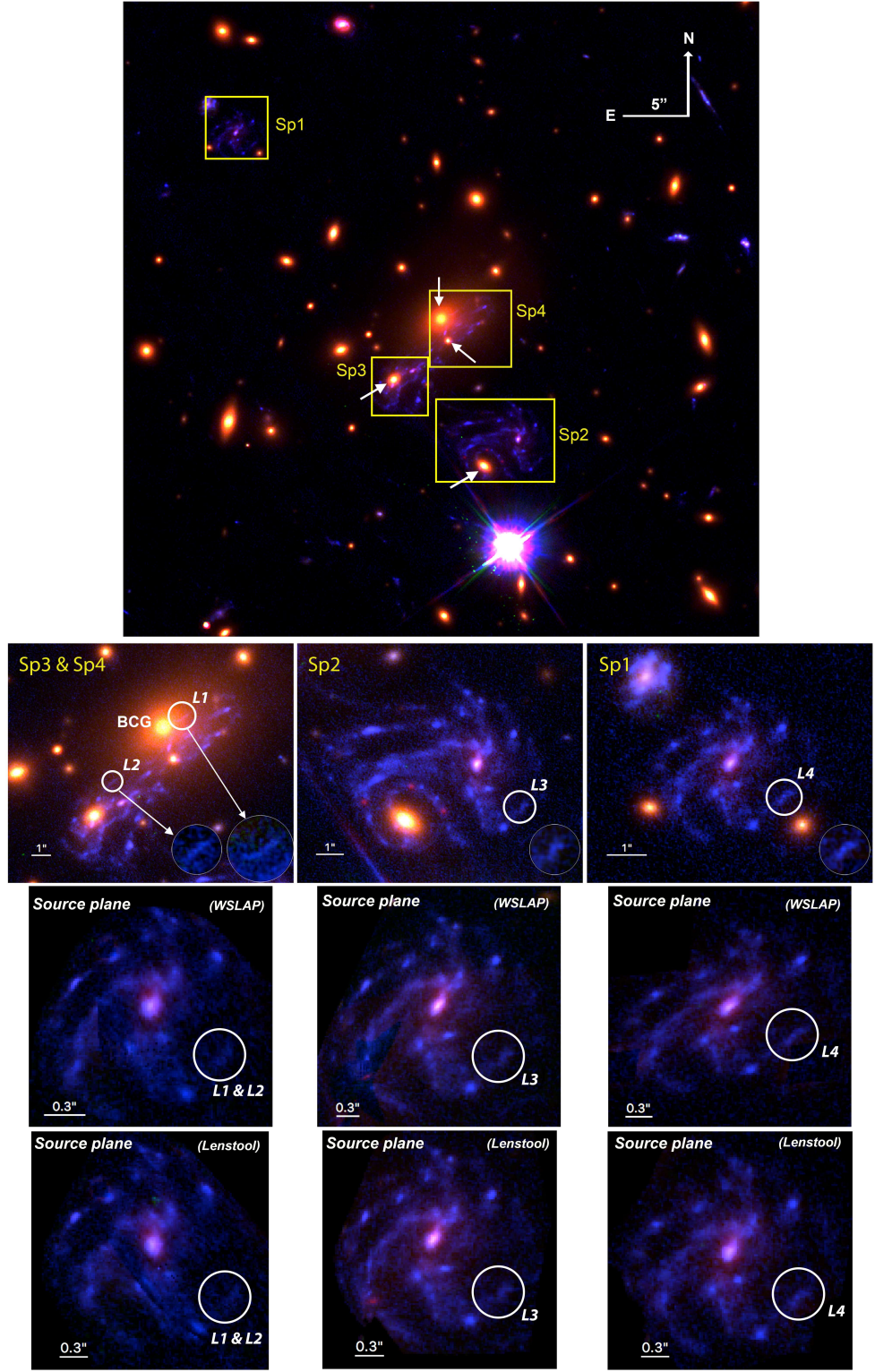}
\caption{\label{fig:DelensImg}{\it Row 1:} HFF data of MACS 1149 ($0\farcm8\times0\farcm8$) in the F125W (red), F814W (green) and F435W (blue) bands. {\it Row 2:} close-up of the individual lensed images Sp1-4. Sp4 is projected close to the BCG, with an HII region in its spiral arm labelled L1 projected close to the BCG center. This HII region appears four times, labelled L2-4 in the other counter images, and is shown in greater detail by a magnified insert at the bottom right corner of each panel. {\it Row 3 \& 4:} the delensed images of the spiral galaxy obtained by applying a free-form WSLAP+ lens model and a parametric \textit{Lenstool} model to each of the multiple images, respectively. Sp3 and Sp4 are combined to form one delensed image as Sp4 is lensed from only a small segment of the entire spiral galaxy. The three independent source plane images of this spiral galaxy are in good agreement with each other, even though Sp1 is relatively far from the cluster center where the lens model is less well constrained. }
\end{figure*}

Figure \ref{fig:DelensImg} shows a $0.\arcmin8\times0.\arcmin8$ multi-band image of MACS 1149 constructed from the F125W (red), F814W (green) and F435W (blue) filters of the HFF. The most prominent lensed image is that of a multiply-lensed spiral galaxy at $z=1.4888$. With the lensed images magnified by up to $\sim200$ times, this object is one of the most highly magnified lensed galaxies yet discovered \citep{Zitrin&Broadhurst2009_MACS1149}. We label the individual multiply-lensed images as Sp1-4, as shown in Figure \ref{fig:DelensImg} (row 2). Sp1-3 correspond to contain complete images of this spiral galaxy, whereas Sp4 corresponds to only a partial image; note that certain features in both Sp3 and Sp4 appear more than once owing to additional lensing by the BCG and the bright elliptical galaxy near Sp3 (indicated by the white arrow closest to this lensed image). One of the numerous HII regions in Sp4, labelled ``L1'' and which lies closest to the BCG, has a curved (banana-like) shape. As we will show in section \ref{lens_modelling}, L1 is itself a doubly-lensed image with a critical curve passing through it as predicted by different constructions of lens models for this cluster. The radius of curvature of L1 is only $\sim0\farcs6$, which as we will show is much smaller than the Einstein radius of either the BCG or the cluster and therefore requires a compact lensing mass that is close to L1 on the sky. 

To better reveal L1 as well as nearby lensed features at the vicinity of the BCG centre, we also subtracted the BCG light from the F435W data, as shown in Figure \ref{fig:WSLAPoutput_BCGcompare}, 1st panel. The subtraction is not model dependent given the uniformity of the colour of early type galaxies, as we can simply scale the very bright image of the BCG at F160W band and subtract it from the F435W band. The scaling factor is decided by trial and error so as to best remove the BCG light without over-subtraction. This removal is straightforward in the F435W band as the BCG is barely detectable and lensed images from the background spiral galaxy at $z=1.4888$ have the highest contrast against the BCG light in this passband. As can be seen in Figure \ref{fig:WSLAPoutput_BCGcompare}, 1st panel, only noise remains after subtraction. 

\subsection{MUSE/VLT}
We analysed the data of MACS 1149 taken with Multi-Unit Spectroscopic Explorer (MUSE) at the Very Large Telescope (VLT). Sp1-4 was observed for the purpose of studying the kinematics of the background spiral galaxy, which assists the strong lensing analysis. We retrieved the ``phase-3'' processed data of program ID 294.A-5032 (PI: C. Grillo) from the ESO Science Archive Facility\footnote{\url{http://archive.eso.org/cms/data-portal.html}}, and used the ``MUSE-DEEP'' product, where all individual exposures taken on different days under the same program ID are fully calibrated (with MUSE pipeline muse-1.6.1) and combined. This data set was originally presented in \cite{Grillo2016}, and all details of the observations are described there.

The MUSE instrument was configured for its wide field of view (FOV) ($1\arcmin\times1\arcmin$) mode without the AO system. The standard MUSE IFU and spectrograph configuration provides a spectral cube covering a wide wavelength range (4750-9350\AA) with a spectral resolution of $R\sim3000$ at a pixel scale of $0\farcs2\times0\farcs2$. The total exposure time was 17280 s, and the seeing size was $1\farcs0$ based on the measurement of a bright star within the FOV on the combined frame at a wavelength near the redshifted [OII] emission ($\sim9300$\AA; see section \ref{MUSE}). We refined the astrometry by aligning the stellar continuum map generated from our spectral fitting analysis with the HST images to an accuracy of $<0\farcs1$.


\section{Lens modelling}\label{lens_modelling}

\subsection{WSLAP+ model of MACS 1149}
The global lens model of the cluster is derived using our code  WSLAP+ (Weak and Strong Lensing Analysis Package $+$) \citep{Diego2005,Diego2007,Ponente2011,Sendra2014}. This method adopts a free-form philosophy where the lens plane is divided into a pixelated grid. Each pixel is represented as a Gaussian mass profile, where the FWHM may be varied to generate a multi-resolution grid, or is held constant to provide a uniform grid \citep{Diego2005}. The division of the lens plane into grid points allows us to divide the deflection field, $\alpha$, into the individual contributions to the deflection field from the pixel grid. For MACS 1149, we used a multi-resolution grid with 280 cells in total. The resolution of the grid in the cluster centre is $2\farcs16$, and the resolution in the edges is $6\farcs56$. A further improvement was implemented by including member galaxies of the cluster with NFW mass profiles scaled by their measured luminosities \citep{Sendra2014}, and for which the only free parameter is the scaling of the M/L ratio (size scale), for all bright member galaxies included in the model. This M/L ratio and the Gaussian masses in the grid points are derived by means of minimising a quadratic function. The minimum of this quadratic function is also the solution of a system of linear equations that describe the observed data (see Eq.\ref{eq_lens_system} below). More specifically, our method is described below.

Given the standard lens equation, 
\begin{equation} \beta = \theta -
\alpha(\theta,\Sigma(\theta)), \label{eq_lens} 
\end{equation} 
where $\theta$ is the observed position of the source, $\alpha$ is the deflection angle, $\Sigma(\theta)$ is the projected surface mass density of the cluster at the position $\theta$, and $\beta$ is the position of the background source. Both the strong lensing and weak lensing observables can be expressed in terms of derivatives of the lensing potential, 
\begin{equation} 
\psi(\theta) = \frac{4 G D_{l}D_{ls}}{c^2 D_{s}} \int d^2\theta'
\Sigma(\theta')ln(|\theta - \theta'|), \label{eq_psi} 
\end{equation}
where $D_l$, $D_{ls}$ and $D_s$ are the angular diameter distances to the lens, from the lens to the source and from the observer to the source, respectively. The unknowns of the lensing problem are in general the surface mass density (or masses in our grid points) and the positions of the background sources. The weak and strong lensing problem can be expressed as a system of linear equations that can be represented in a compact form \citep{Diego2007}, 
\begin{equation}
\Theta = \Gamma X, \label{eq_lens_system} 
\end{equation} 
where the measured strong and weak lensing observables are contained in the
array $\Theta$ of dimension $N_{\Theta }=2N_{SL} + 2N_{WL}$, the
unknown surface mass density and source positions are in the array $X$
of dimension $N_X=N_c + N_g + 2N_s$, and the matrix $\Gamma$ is known
(for a given grid configuration and fiducial galaxy deflection field, 
see below) and has dimension $N_{\Theta }\times N_X$.  $N_{SL}$ is the number
of strong lensing observables (each one contributing with two constraints,
$x$, and $y$), $N_{WL}$ is the number of weak lensing observables
(each one contributing with two constraints, $\gamma_1$, and $\gamma_2$), and $N_c$ is the number of grid points (or cells) that we use to divide
the field of view. $N_g$ is the number of deflection fields (from
cluster members) that we consider.  $N_s$ is the number of background
sources (each contributes with two unknowns \citep{Sendra2014}, $\beta_x$, and $\beta_y$. The solution is found after
minimizing a quadratic function that estimates the solution of the
system of Eq. \ref{eq_lens_system}. For this minimization we
use a quadratic algorithm which is optimized for solutions with the
constraint that the solution, $X$, must be positive \citep{Diego2005}. This is particularly important since by imposing this constraint we avoid the
unphysical situation where the masses associated to the galaxies are
negative (that could, from the formal mathematical point of view, otherwise provide a reasonable solution to the system of linear Eq. \ref{eq_lens_system}). Imposing the constrain $X>0$ also helps in regularizing 
the solution as it avoids large negative and positive contiguous fluctuations. For modelling MACS 1149, we did not use weak lensing constraints. Strong lensing constraints come from 16 multiply-lensed galaxies, some of which contain individual resolved features that add up to the total number of constraints. In the Appendix, we mark the locations of multiply-lensed galaxies in Figure \ref{fig:image_ID} and list the coordinates of multiple images used as constraints in Table \ref{image-constraints}.

Previous work has shown how the addition of the small deflection fields from member galaxies can help improve the mass  determination when enough constraints are available \citep{Kassiola1992,Kneib1996,Sendra2014}. In a previous study, we quantified via simulations how the addition of deflections from all the main member galaxies helps improve the mass reconstruction with respect to our previous standard  non-parametric method \citep{Sendra2014}. Strongly lensed galaxies are often locally affected by member galaxies. However, these perturbations cannot be recovered
in grid based reconstructions because the lensing information is too sparse to resolve member galaxies. 

For our study we select elliptical galaxies in the cluster and assign a mass according to luminosity. For the fiducial deflection field we assume that the mass of the member galaxies scales by a fixed M/L ratio. The optimization procedure determines the proportionality constant that allows for the best reproduction of the data. As mentioned above, for the mass profiles we assume a NFW profile \citep{NFW1996}, and adopt a self-similarity so that the scale radius is proportional to mass. Note that the choice of the particular profile for these perturbing galaxies is not very relevant in terms of reproducing the multiply-lensed images of a given background galaxy, as the deflection angle is small compared to that produced by the smooth cluster component. We use two deflection fields for the cluster members (i.e $N_g=2$, see definition of $N_g$ above), thus allowing for different M/L ratios for the separate deflection fields. The first one is associated to the BCG and three other galaxies that are the most close to the lensed images of the background spiral galaxy of interest (pointed with the white arrows in Figure \ref{fig:DelensImg}), and the second one contains the deflection field from the remaining dominant galaxies in the cluster. Each deflection field contributes in our model as one free parameter (its amplitude with respect to the fiducial amplitude). More details of the global lens modelling of this cluster can be found in our previous study \cite{Diego1149}. This model successfully predicted the reappearance of the SN Refsdal \citep{2016ApJ...817...60T}.

\subsection{Fine-tuning BCG mass model}\label{finetune_BCG}
\begin{figure*}[tp!]
\centering
\graphicspath{{/Users/Mandy/GoogleDrive/BH_paper/ApJ/figures/}}
\includegraphics[width=14cm]{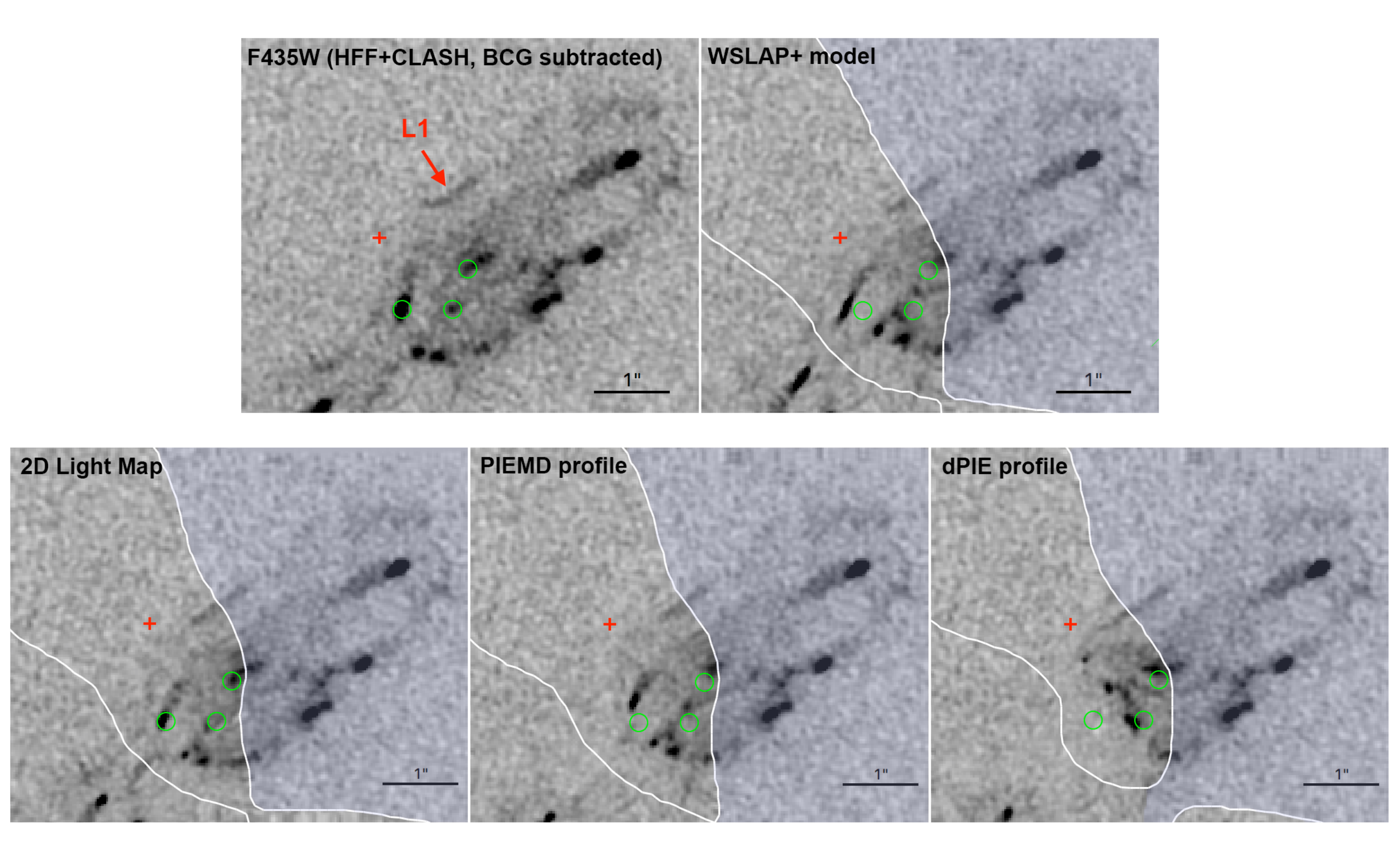}
\caption{\label{fig:WSLAPoutput_BCGcompare} \textbf{The red cross indicates the light centroid of the BCG.} {\it Upper left:} the data in the region of interest close to the centre of the BCG (after the BCG light is subtracted). {\it Upper right:} the relensed image from our free-form lens model (WSLAP+) where a NFW mass profile is assumed for the BCG. The blue-highlighted region on the right-hand side of a critical curve (white line) is the input data used for relensing, and that on the left-hand side the corresponding relensed images. Note that image L1 is predicted to be more straight than is actually observed. {\it Lower row:} same as upper right, but differs in the way the BCG mass distribution is modelled to seek closer agreement between the predicted and observed relensed images. By using the 2D light distribution for the BCG, we obtain a better agreement in terms of the reproduction of the relensed images (lower left), but again with the obvious exception of the curved appearance of L1. In the lower middle and right panels, we model the BCG with a dPIE and a PIEMD mass profile, as adopted by \cite{Grillo2016}. The green circles point out three major features close to the centre of the BCG in the relensed region, and they are plotted in all the model panels to indicate the offset of the model predicted positions. The mean rms offset of these 3 relensed images from their positions in the data ($<rms_\textit{i}>$) in the 4 models are: $0\farcs22$, $0\farcs13$, $0\farcs21$, and $0\farcs49$.}
\end{figure*}

In Figure \ref{fig:WSLAPoutput_BCGcompare} (upper row), we show the critical curves in the region around the BCG as predicted by lens model produced using WSLAP+. Owing to the a local perturbation to the lensing associated with the BCG, lensed images within $\sim6.4$ kpc radius from the BCG centre are triply lensed. We can therefore delens (i.e., trace the lensed feature back to the source plane with the deflections predicted by the lens model) one set of the multiple images back to the source plane, and relens (i.e., re-trace the source to the image plane with the deflections predicted by the lens model) them back to the image plane to reproduce the other two sets of images in this region. The detailed features of the images thus reproduced serve as local constraints to the BCG mass distribution. Figure \ref{fig:WSLAPoutput_BCGcompare} shows that although the matching between the actual images and those reproduced through delensing and then relensing is reasonably good, there is clearly room for improvement by locally adjusting the BCG mass distribution. Another point to notice is that L1 is predicted to be straight in the WSLAP+ solution. The same straight appearance for L1 is found in the cluster lens model produced by \cite{Grillo2016} using the parametric algorithm GLEE \citep{Suyu2012}, as we will discuss in section \ref{consistency_checks}.  In section \ref{lenstool_section}, we show that the cluster lens model we independently produce using \textit{Lenstool} \citep{lenstool} also predicts L1 to be straight.

To improve the mass model of the BCG, we tried three different profiles for its projected two dimensional mass distribution (while keeping the rest of the WSLAP+ solution fixed, as is necessary to preserve the good global agreement obtained for all the multiply-lensed images with this solution). The first simply assumes that the projected two-dimensional mass distribution of the BCG follows its projected two-dimensional light distribution, with the only parameter being the normalisation factor (i.e., M/L ratio). The second assumes a dual pseudoisothermal elliptical mass distribution (dPIE) \citep{dPIE2007} with the convergence given by
\begin{equation}
\kappa = \frac{\theta_{E}}{2}\left (\frac{1}{\sqrt{R_{\epsilon}^2+r_{c}^2}}-\frac{1}{\sqrt{R_{\epsilon}^2+r_{t}^2}}\right) , \label{eq_dPIE} 
\end{equation}
where $r_{c}$ is the core radius and $r_{t}$ is the truncation radius. The third assumes a pseudo-isothermal elliptical mass distributions (PIEMD) \citep{PIEMD1993}, with the convergence given by
\begin{equation}
\kappa = \frac{\theta_{E}}{2\sqrt{R_{\epsilon}^{2}+r_{c}^{2}}} , \label{eq_PIEMD} 
\end{equation}
where $r_{c}$ is the core radius. In both dPIE and PIEMD profiles, $R_{\epsilon}$ is defined as
\begin{equation}
R_{\epsilon} = \frac{x^2}{(1+\epsilon)^2}+\frac{y^2}{(1-\epsilon)^2} , \label{eq_elliptical radius} 
\end{equation}
where $\epsilon$ is the ellipticity of the profile, and the position angle is fixed to be at the observed position angle of the BCG's stellar component.

In Figure \ref{fig:WSLAPoutput_BCGcompare} (lower row), we show the relensed reproductions from three different models for the BCG mass distribution as described above. In both the dPIE and PIEMD profiles, we began with an initial zero $\epsilon$. For the dPIE profile, we used $r_{c} = 0\farcs24$, which corresponds to the core radius in the light profile of the BCG, and $r_{t} = 1\farcs5$, which is close to the effective radius of the BCG. For the PIEMD profile, we also used $r_{c} = 0\farcs24$. We found that by changing the parameters ($\epsilon$, $r_{c}$ and $r_{t}$) in the dPIE and PIEMD mass profiles to improve the agreement between the predicted and observed relensed images, the mass distribution approached the two-dimensional light distribution of the BCG. Hence in all the following steps, we use the lens model where the BCG mass distribution is represented by its two-dimensional light distribution. With the best-fit normalisation factor, the mass contributed from the BCG scaled with the two-dimensional light distribution is $6.3\times10^{11}M_\odot$ within a cylinder of $r<30\textrm{kpc}$, while the total projected mass within the same region is $7.1\times10^{12}M_\odot$.

At this point, we would like to emphasize that both prior to and after improving the mass distribution of the BCG, L1 is predicted to be straight. This result reflects the fact that, to curve L1 on a radius of curvature of $\sim0\farcs6$, a compact local deflector is needed as described in sections \ref{result_section} and \ref{lenstool_section}.


\section{Detailed investigation of L1}\label{L1_investigation_section}
\subsection{Spectroscopic analysis using MUSE data}\label{MUSE}
\begin{figure*}[tp!]
\centering
\graphicspath{{/Users/Mandy/GoogleDrive/BH_paper/ApJ/figures/}}
\includegraphics[width=16cm]{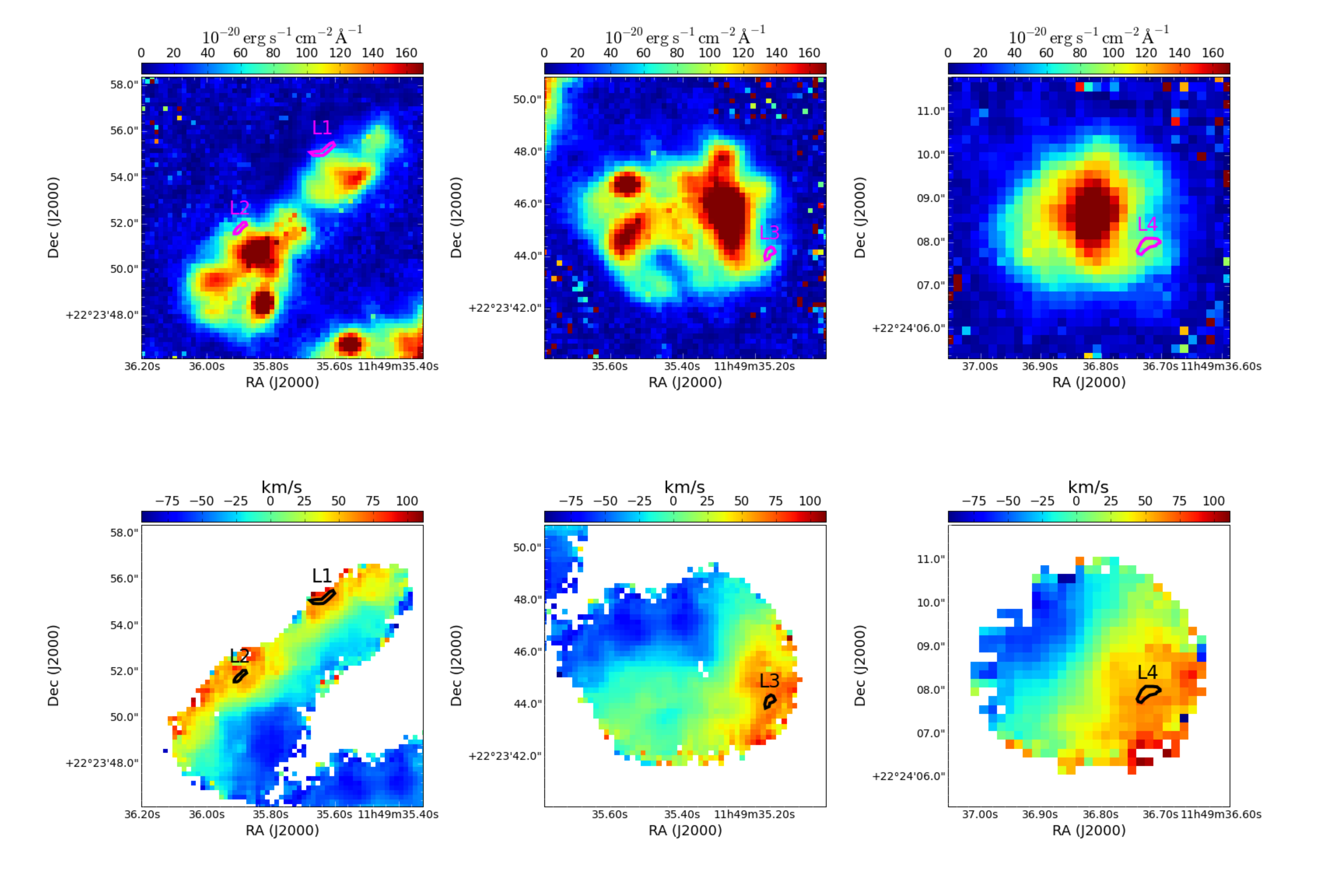}
\caption{\label{fig:velocity_map}[OII] flux map and velocity map at the rest frame of the spiral galaxy derived from the [OII] line. {\it Upper row:} the isophotal contours indicating the positions of L1-4 are plotted on the [OII] flux map. {\it Lower row:} L1-4 belong to the same kinematic region of the background spiral galaxy, confirming the multiple image identifications.}
\end{figure*}

\begin{figure*}[tp!]
\centering
\graphicspath{{/Users/Mandy/GoogleDrive/BH_paper/ApJ/figures/}}
\includegraphics[width=15cm]{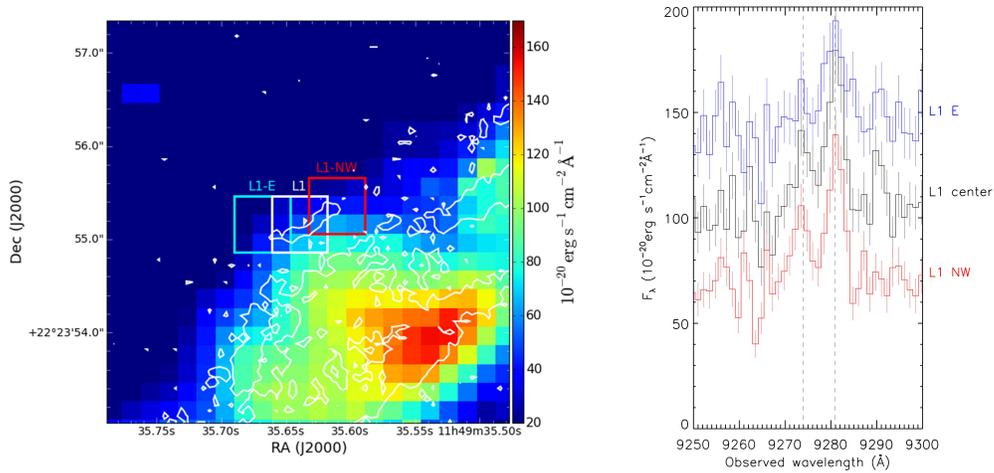}
\caption{\label{fig:l1_profile_aperture}Spectral profile of L1 derived from the [OII] line. {\it Left:} the spectral aperture on top of the [OII] flux map. White contours are the 0.0015 electron/s isophotal contours from the F435W HFF data image. {\it Right:} the one-dimensional spectral profiles taken from the three apertures indicated in the left panel. The vertical dashed lines are the rest-frame [OII] doublet lines. The agreement of the spectral profiles from the three apertures spanning the entire L1 indicates that the whole arc of L1 belongs to the background spiral galaxy.}
\end{figure*}

\begin{figure}[tp!]
\centering
\graphicspath{{/Users/Mandy/GoogleDrive/BH_paper/ApJ/figures/}}
\includegraphics[width=\linewidth]{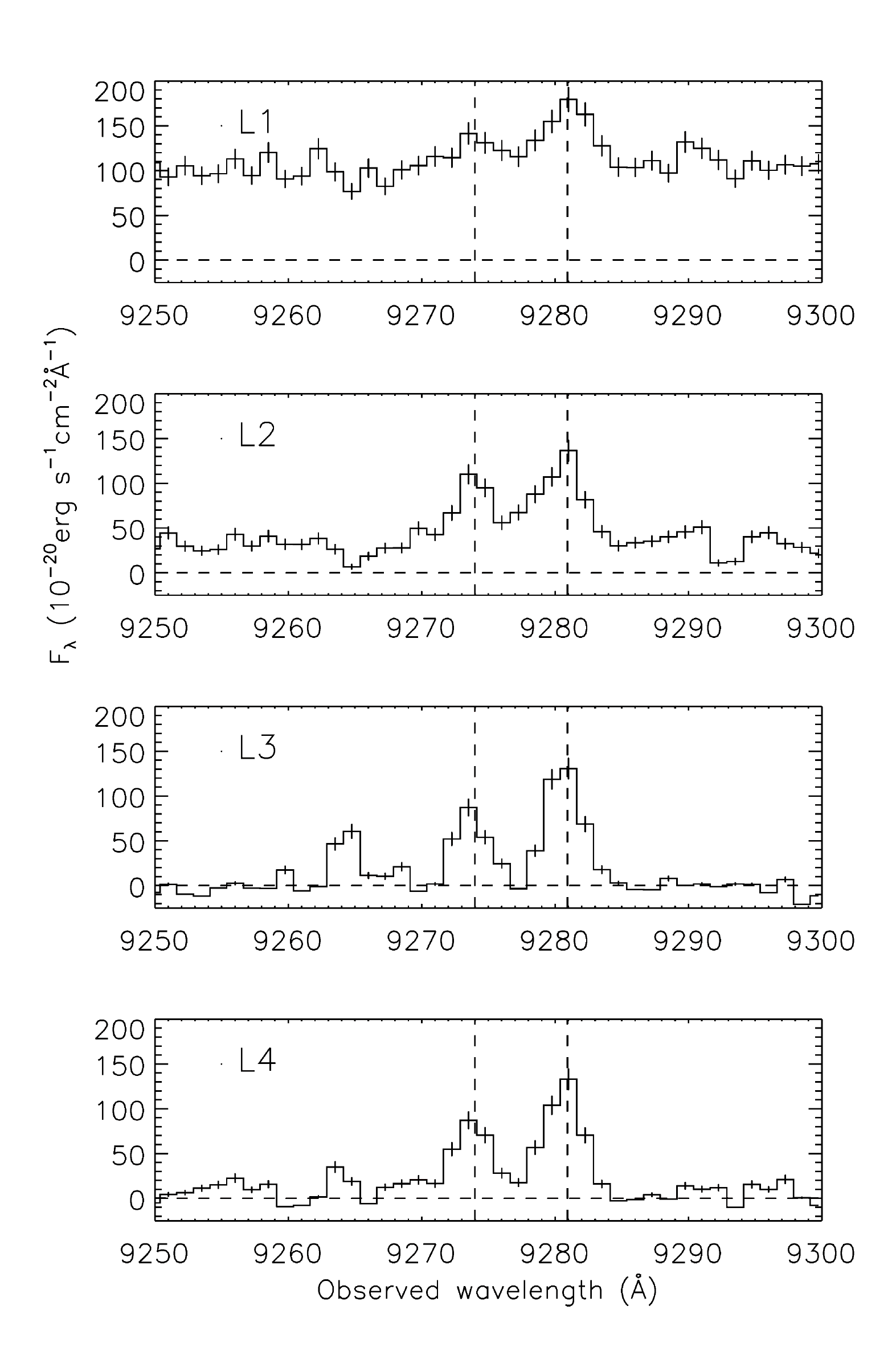}
\caption{\label{fig:profile_l1-4}Spectral profile of L1-4 derived from the [OII] line. The vertical dashed lines are the rest-frame [OII] doublet lines. The line widths at the position of L1 and L2 are bigger that that at the position of L3 and L4 due to the large shear induced by lensing.}
\end{figure}

To ensure that L1 is a single contiguous feature and that its lensed counterparts have been correctly identified, we extracted the [OII] $\lambda\lambda$ 3726, 3729 doublet emission of the background lensed spiral galaxy using simple spectral fitting. This doublet is the only emission line feature expected for star-forming galaxies at z=1.4888 \citep{Grillo2016} within the MUSE spectral coverage. We modelled the [OII] doublet as two Gaussian functions having the same linewidth at the same redshift but having different amplitudes, and fitted these functions to the spectra at 9200-9300\AA  (where sky emission lines are weak) spanning the [OII] doublet, where sky emission lines are weak. Around L1 where stellar light from the BCG makes the spectrum relatively bright, stellar absorption lines are seen. We extracted the spectrum of the BCG at its north-eastern region where no contamination by [OII] from the background spiral galaxy is evident, and used it as a template for the BCG stellar spectrum in the fit. In the fitting, we summed spectra over $3\times3$ pixels ($0\farcs6\times0\farcs6$) to improve the S/N without losing the seeing-limited ($1\farcs0$ at FWHM) spatial resolution. The free parameters of the fit are redshift, line width, flux ratio of the doublet, and scaling factor of the BCG stellar light. We utilised a non-linear least squares curve fitting library MPFIT \citep{Markwardt2009} for the fit. From the results of the fit, we generated the velocity field map for Sp1-4 as shown in Figure \ref{fig:velocity_map}, the detailed local spectral profile for L1 as shown in Figure \ref{fig:l1_profile_aperture}, and the one-dimensional spectral profiles for L1-4, as shown in Figure \ref{fig:profile_l1-4}. L1-4 show similar spectral profiles and correspond to the same kinematic region in the lensed background spiral galaxy, in agreement with our multiple image identification. The consistent spectral profiles of L1 taken from the three apertures shown in Figure \ref{fig:l1_profile_aperture} (left panel) confirm that the entire arc of L1 belongs to the same kinematic region in the lensed background spiral galaxy, as our model predicts.


\subsection{Examine the curvature of L1 using all multiply-lensed images}\label{consistency_checks}

\begin{figure*}[tp!]
\centering
\graphicspath{{/Users/Mandy/GoogleDrive/BH_paper/ApJ/figures/}}
\includegraphics[width=13cm]{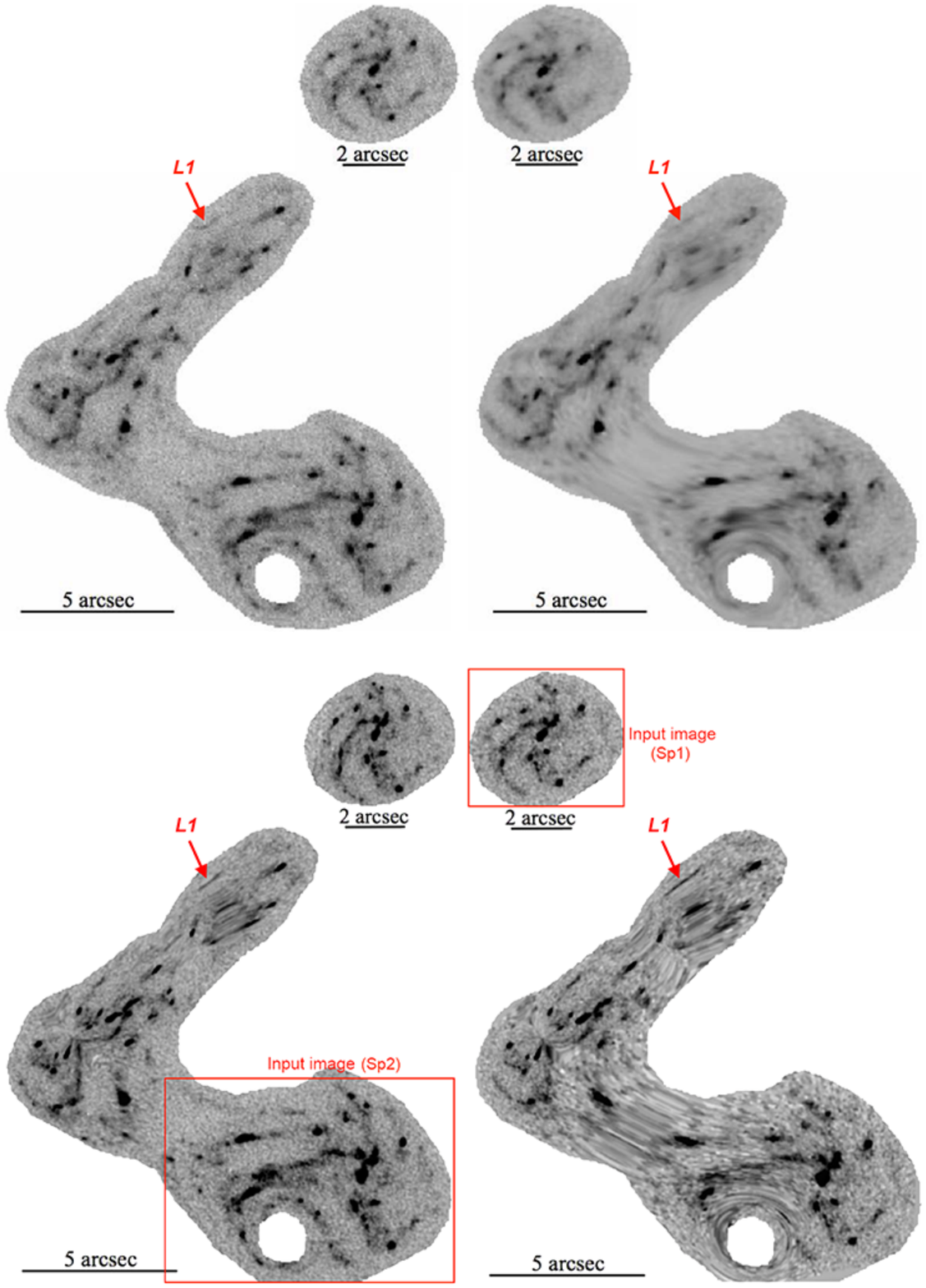}
\caption{\label{fig:largefield}{\it Upper row:} Figure 7 of \cite{Grillo2016}. The left-hand side is the original data but with cluster members subtracted to better reveal the multiply-lensed spiral galaxy. Notice that L1 appears curved in the subtracted data, similar to its appearance in our subtracted image. Right-hand side is the model-generated image from \cite{Grillo2016}, showing good general agreement with the data on the left. Notice that L1 is predicted to be straight in their model, as in our work when we do not impose an extra deflector in the BCG. {\it Lower row:} predicted relensed images of the spiral galaxy obtained by relensing Sp2 in the left panel and Sp1 in the right panel, utilizing our lens model from WSLAP+ solution and the 2D light distribution of the BCG for its mass profile. In each case, there is good general agreement with the data, except that L1 is straight rather than being curved as is observed.}
\end{figure*}

\begin{figure*}[tp!]
\centering
\graphicspath{{/Users/Mandy/GoogleDrive/BH_paper/ApJ/figures/}}
\includegraphics[width=15cm]{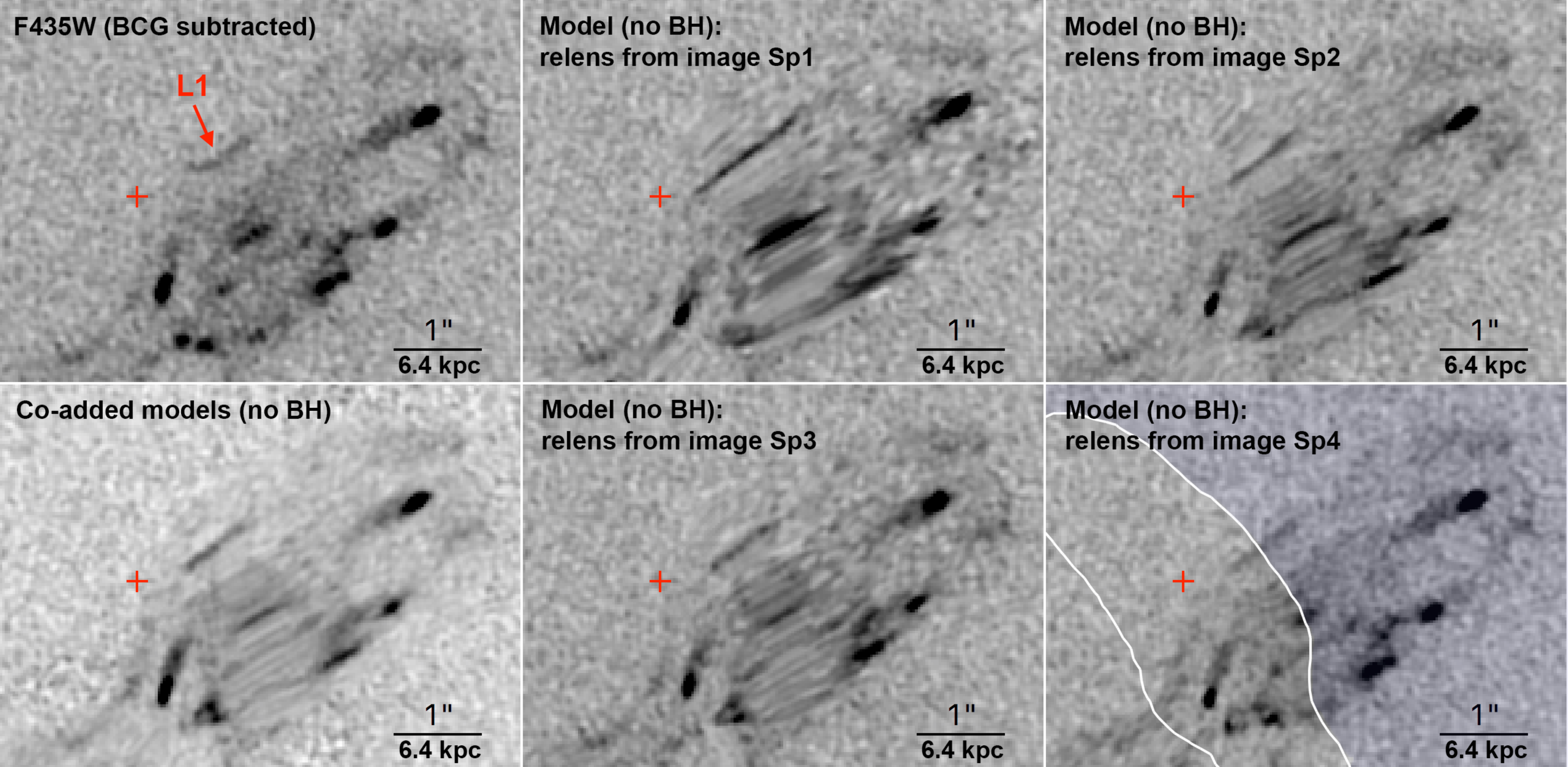}
\caption{\label{fig:consistency_nobh}Detailed comparison of lensed features around the BCG based on relensing of four independent lensed images of the background spiral galaxy (Sp1-4, see Figure \ref{fig:DelensImg} for the locations of theses images in the cluster). {\it Upper left:} image in the F435W band. The relensed images in the remaining panels all assume the same BCG mass distribution, based on its 2D light map as described in section \ref{finetune_BCG}. The light centroid is marked with a red cross for reference in all the panels. Note that in the lower right panel, only the blue-highlighted region on the right hand side is used as the input for relensing. The consistency between the relensed images and their actual appearance is good considering the differences in location and magnifications of each of the input data images for relensing. The obvious exception is the curvature observed in L1, not predicted in this model, underscoring the need for additional deflection local to L1.}
\end{figure*}

We examine the need for a local deflector to produce the curvature of L1 by delensing each of the four spiral galaxy images (Sp1-4) back to the source plane and then making relensed projections to the image plane (see Figure \ref{fig:DelensImg} for the locations of Sp1-4 in the cluster). In this way, we can also determine the level of agreement between these independently produced images. This agreement check is useful for establishing confidence in the overall accuracy of our cluster lens model. Of course, we are particularly interested in the predicted appearance of L1, when delensed and then relensed from its different counterparts (L2-L4), and therefore investigate the level of agreement for this feature between the different lensed images.

The lens model on the large scale is generally very accurate at the level of 3\% in terms of the deflection angle, which can be appreciated by delensing any lensed image and relensing it to form the counter images well separated in angle. In Figure \ref{fig:largefield}, we show the agreement between our WSLAP+ free-form model and the parametric model of \cite{Grillo2016}, both of which successfully predicted the position and time of the reappearance of SN Refsdal in one of the lensed images of the background spiral galaxy. In Figure \ref{fig:largefield} (lower row), we predict the large scale distribution of the lensed images of the spiral galaxy by delensing and then relensing, alternatively, the lensed images labelled ``Input image (Sp1)" and ``Input image (Sp2)". The level of agreement that we find is comparable with the general level of relensing accuracy determined for this cluster by other independent work \citep{Zitrin&Broadhurst2009_MACS1149,Diego1149,Treu2016}, and also for other complex clusters with deep Hubble data \citep{broadhurstetal2005,Halkola2008}. Figure \ref{fig:largefield} deliberately follows the published image format from \cite{Grillo2016} to make this comparison as clear as possible. Very good agreement is seen between our free-form method and their parametric approach to lens modelling.

When delensing and then relensing Sp1-4, the agreement between counter images is better for more closely separated images as is expected. A small systematic shift of $1\farcs1$ and $0\farcs3$ were applied to the more distant counter-images Sp1 and Sp2, and less than $0\farcs1$ to Sp3 so that they become better aligned. These shifts show an accuracy limitation at the large scale. The corresponding adjustments corrects the systematic offset between the predictions based on different relensed images and the data in the central region of the lensed field, thus allowing us to make a detailed comparison of the internal features of the spiral galaxy between each relensed images and the region of interest around the BCG.We emphasize at this point that these small systematic errors reflect the accuracy of the cluster-scale lens model given the constraints available from all the multiply-lensed background galaxies, and that no  modification to this cluster model can produce the curved appearance of L1 with a radius of curvature of $\sim0\farcs6$. The large and smooth deflection angles that the cluster-scale lensing induces are $\sim30\arcsec$ in size, and any modification of the cluster deflection field can therefore only shift a small image uniformly without bending it. 

Figure \ref{fig:consistency_nobh} shows delensed and then relensed images in the region around the BCG by delensing and then relensing, alternatively, the lensed images Sp1-4. Note that when relensing Sp4, we only include the part of Sp4 that is indicated by the blue-highlighted region (lower right panel), and therefore only the portion of L1 to the right of the critical curve, to compare its delensed and relensed image to its counter image on the opposite side of the critical curve. We use the best fit solution as described earlier comprising the grid to represent the large scale distribution of dark matter, the 2D light map of the BCG to account for its mass, and a 2D NFW mass distribution associated with each identified member galaxy. As can be seen, all the bright features seen in the data appear in each of the four delensed and then relensed images. To form a deeper delensed and then relensed image, we added together all the individual relensed images as shown in Figure \ref{fig:consistency_nobh} (lower left panel). This deeper image is now noticeably blurred with respect to the individual lensed images and with respect to the data, reflecting the small residual level of difference between the lens mapping and the four independent source images. The purpose of this additional image is to demonstrate that these effects are small, being not much larger than the angular resolution of the data. It does not affect significantly our subsequent conclusions regarding the shape of L1 and the neighbouring features. A clear conclusion from this comparison is that L1 is straight as predicted by delensing and then relensing all four counter images, by contrast with the actual appearance of L1. Note also that the neighbouring internal details are all well reproduced by the delensed and then relensed images with little evidence of any systematic difference in either shape or orientation. As can be seen in Figure \ref{fig:largefield} (upper row), the straight appearance predicted for L1 when delensing and then relensing its counter images is also apparent in the independent lens model presented by \cite{Grillo2016}.


\subsection{Shear effect at the position of L1}\label{shear_section}
\begin{figure*}[tp!]
\centering
\graphicspath{{/Users/Mandy/GoogleDrive/BH_paper/ApJ/figures/}}
\includegraphics[width=\linewidth]{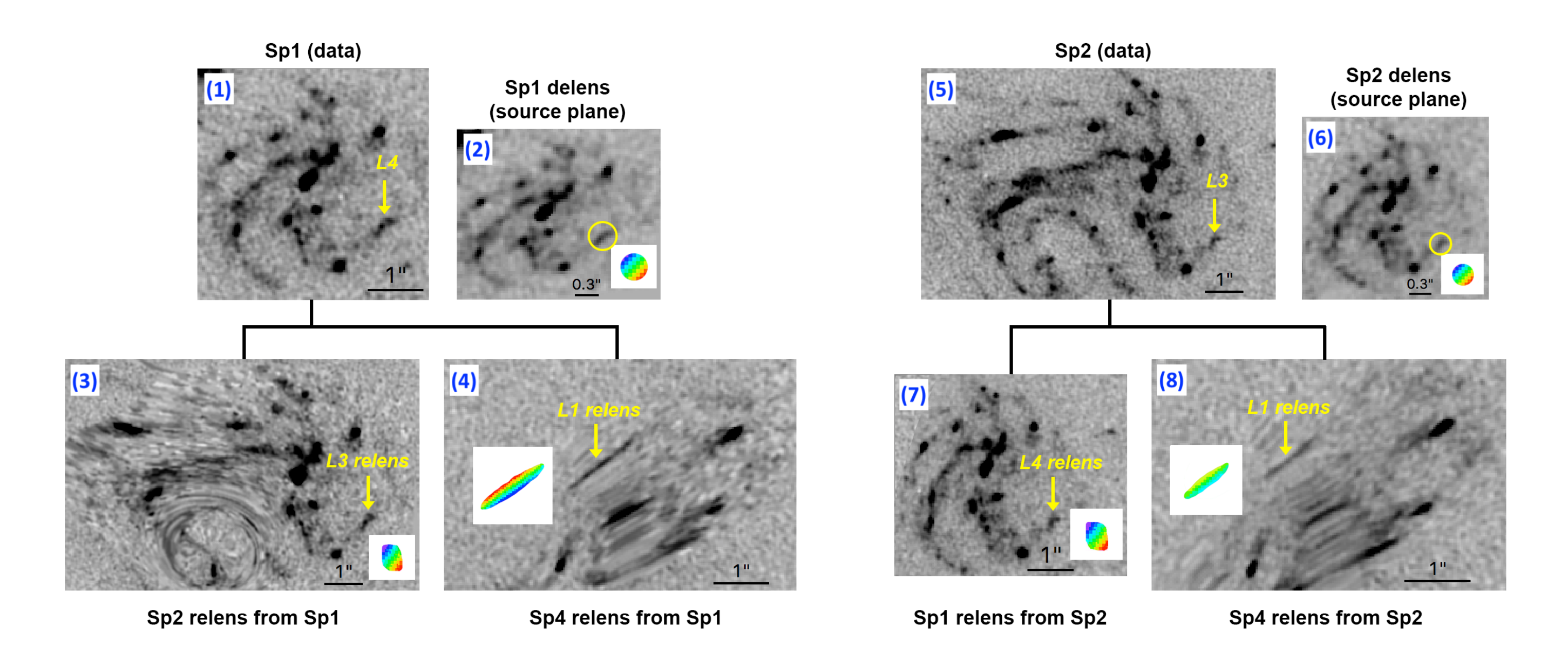}
\caption{\label{fig:shear_illustration}Illustration of shear effect on positions of L1, L3 and L4. {\it (1):} data zooming into the Sp1 region. {\it (2):} delensed Sp1 in the source plane. A color-coded circle as shown in the insert is placed at the position of delensed L4 in the source plane (yellow circle). {\it (3) and (4):} the relensed images of Sp1 at Sp2 and Sp4 region respectively. The projection of the color-coded circle from source plane to the image plane at position of L3 and L1 are shown in the inserts. {\it (5)-(8)} are same as {\it (1)-(4)}, but with Sp2 as the input data for delensing and relensing. This figure demonstrates that the magnitude of shear at position of L1 is much bigger than at the positions of L3 and L4, making L1 more stretched than L3 and L4.}
\end{figure*}

We now demonstrate the shear effect at the position of L1, and contrast it with the shear effect at the positions of L3 and L4. As we will elaborate, this shear difference implies that the bending of L1 cannot be accounted for by intrinsic substructures in the source image.

As shown in Figure \ref{fig:shear_illustration}, we first delensed L3 (contained in Sp1) and L4 (contained in Sp2) to the source plane, then put a color-coded circle at the position of the delensed L3 and L4. We relensed this circle back to the image plane, hence visualising the shear magnitude and orientation of L3 and L4 at the locations of their counter images. As clearly demonstrated in Figure \ref{fig:shear_illustration}, the magnitude of shear at position of L1 is much bigger than the shear at the positions of L3 and L4, making L1 more stretched than L3 and L4. This shear difference can be understood from the fact that the critical curve passes through L1, indicating the gravitational potential gradient being much higher in the vicinity of L1 compared to positions further away from critical curves such as L3 and L4. Therefore, although L1 has intrinsic substructures as observed in L3 and L4, the image formed at the position of L1 is guaranteed to possess a linear shape owing to the high magnitude shear at that particular position. 

An important point to note here is that the small radius of curvature of L1 is only $\sim0\farcs6$. The bending on such a small scale cannot be produced by BCG lenses (which have typical Einstein radii of $\sim5\arcsec$) and cluster lenses (which have Einstein radii of $\sim30\arcsec$ and are responsible for bending the giant arcs). Although an accurate large-scale cluster lens model is essential to account for the lensed images of the spiral galaxy as a whole, the bending of L1 is a local effect that is beyond the influence of the combined lensing from the BCG and the cluster. This fact makes the exploration of the bending imposed on L1 independent of the large-scale lens model (given sufficient accuracy of the cluster lens model as we have demonstrated in the previous sections), as confirmed with a different parametric lens model constructed by \textit{Lenstool} that we will describe in section \ref{lenstool_section}. Furthermore, whatever is responsible for this image curvature should have a deflection field that drops quickly with distance, otherwise the neighbouring images will be noticeably deflected while failing to bend L1 on the small scale observed as we will show in section \ref{extended_deflector_subsection}. In the next section, we will show that a point mass -- a SMBH -- uniquely satisfies these constraints and satisfactorily reproduces the degree of bending seen for L1.


\section{Adding a local deflector}\label{result_section}
\subsection{A point mass}
As we will now show, the curvature of L1 can be reproduced by adding a point mass near the center of the BCG to the lens model. 

\begin{figure*}[tp!]
\centering
\graphicspath{{/Users/Mandy/GoogleDrive/BH_paper/ApJ/figures/}}
\includegraphics[width=15cm]{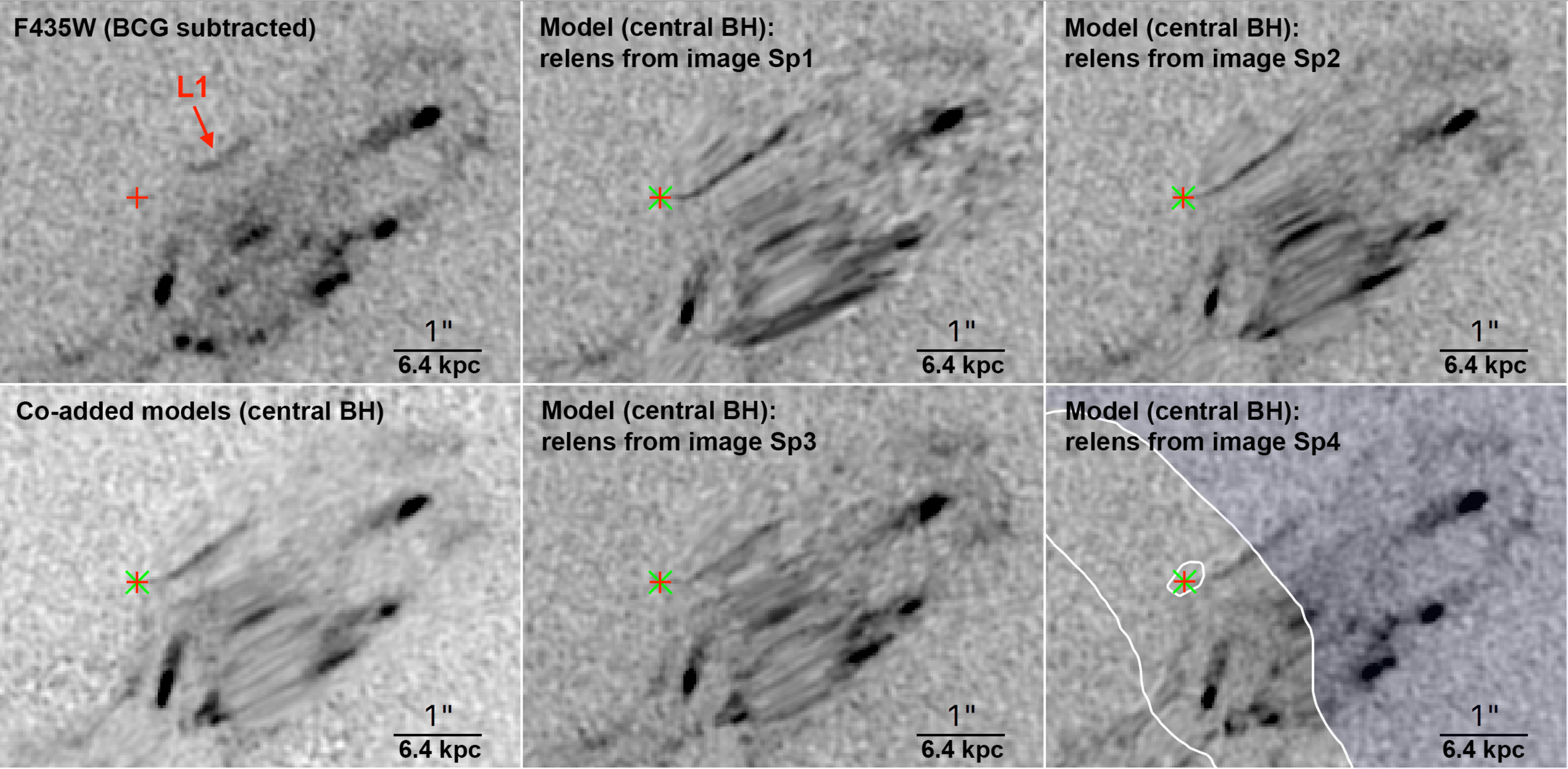}
\caption{\label{fig:consistency_centralbh}Same as Figure \ref{fig:consistency_nobh}, but with the addition of a massive central black hole (green cross) at the location of the light centroid (red cross). The black hole is modelled as a point mass of $8\times10^{9}M_\odot$. The independent predictions and their co-addition clearly shows that adding such a point mass significantly elongates the predicted length of L1 towards the light centre, quite unlike in the actual data where L1 is bent away from the centre.}
\end{figure*}

\begin{figure*}[tp!]
\centering
\graphicspath{{/Users/Mandy/GoogleDrive/BH_paper/ApJ/figures/}}
\includegraphics[width=15cm]{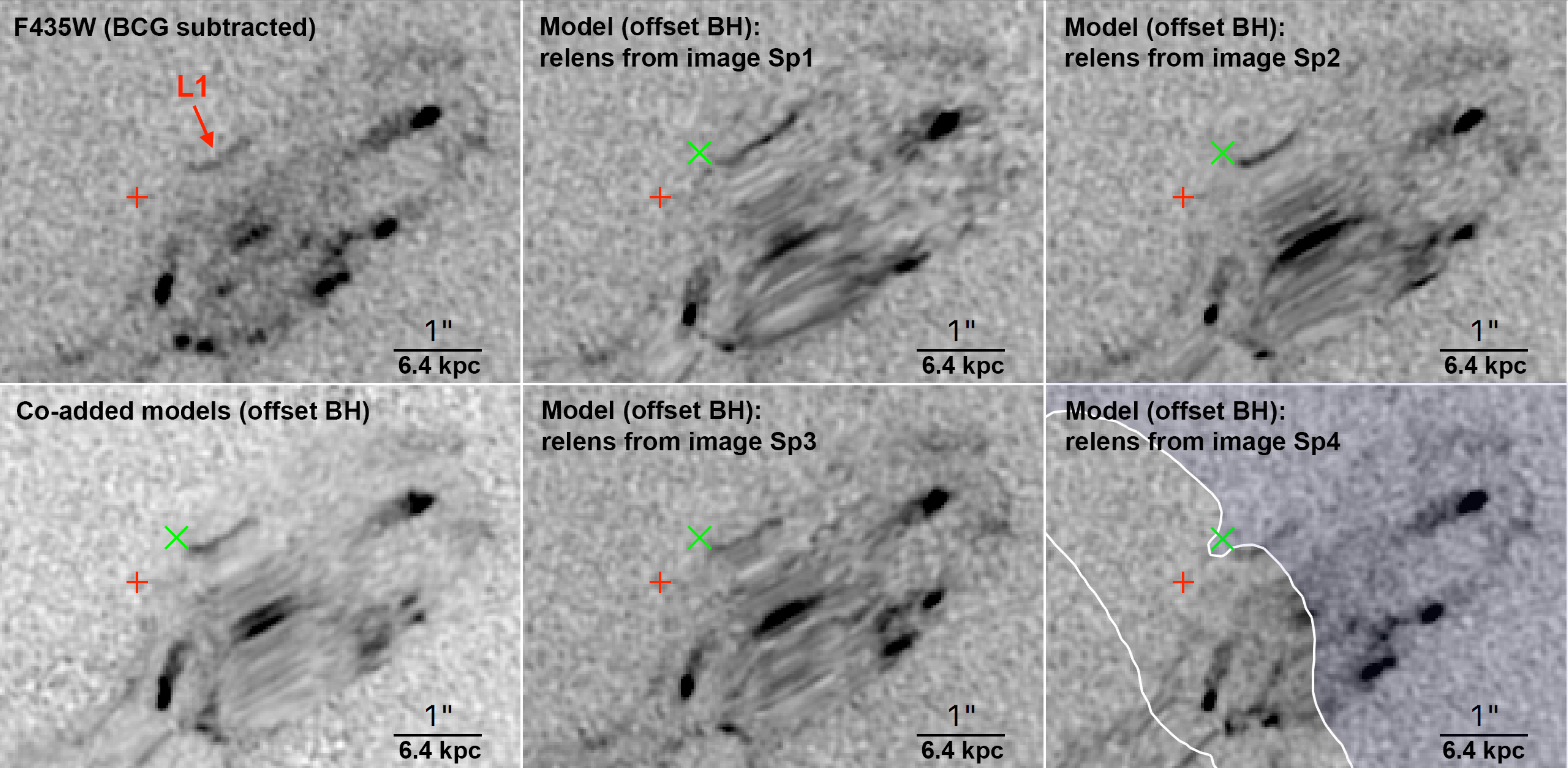}
\caption{\label{fig:consistency_offsetbh}Same as Figure \ref{fig:consistency_centralbh}, but with the addition of a point mass of $8\times10^9M_{\odot}$ at the position of the green cross. Such a point mass introduces a significant curvature in L1, consistent with that seen in the data, without affecting the good fit already obtained for the other images. Note that the BCG mass is scaled down by $\sim6\%$ in this combined best fit, as described in the Supplementary Methods.}
\end{figure*}

\begin{figure*}[tp!]
\centering
\graphicspath{{/Users/Mandy/GoogleDrive/BH_paper/ApJ/figures/}}
\includegraphics[width=15cm]{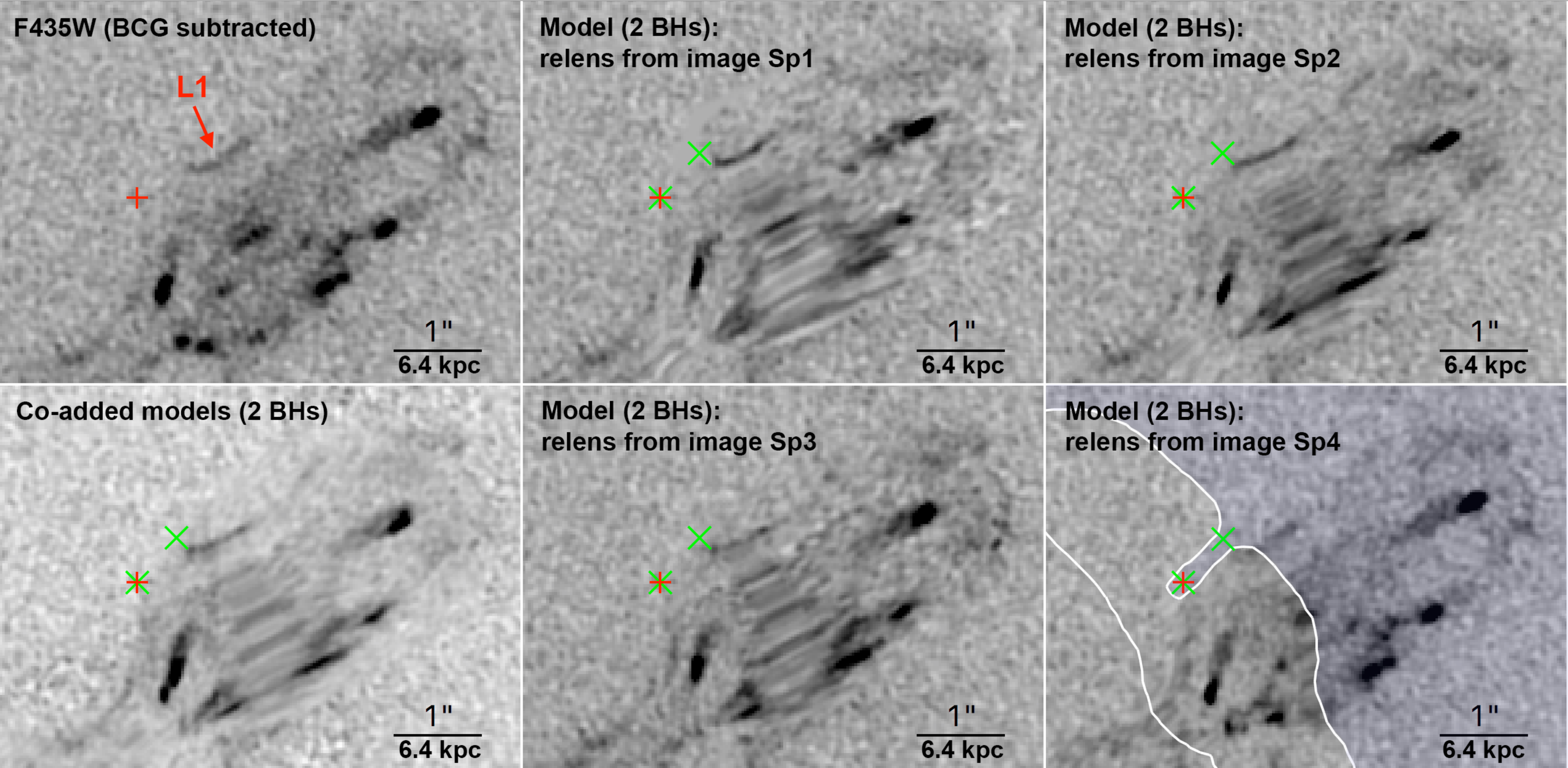}
\caption{\label{fig:consistency_2bh}Same as Figure \ref{fig:consistency_offsetbh}, but but now with the addition of two point masses, one at the centre as in Figure \ref{fig:consistency_centralbh} and the other offset as in Figure \ref{fig:consistency_offsetbh}. This figure shows that there is no need for an additional central black hole, but our model does not exclude its presence.}
\end{figure*}

\begin{figure*}[tp!]
\centering
\graphicspath{{/Users/Mandy/GoogleDrive/BH_paper/ApJ/figures/}}
\includegraphics[width=15cm]{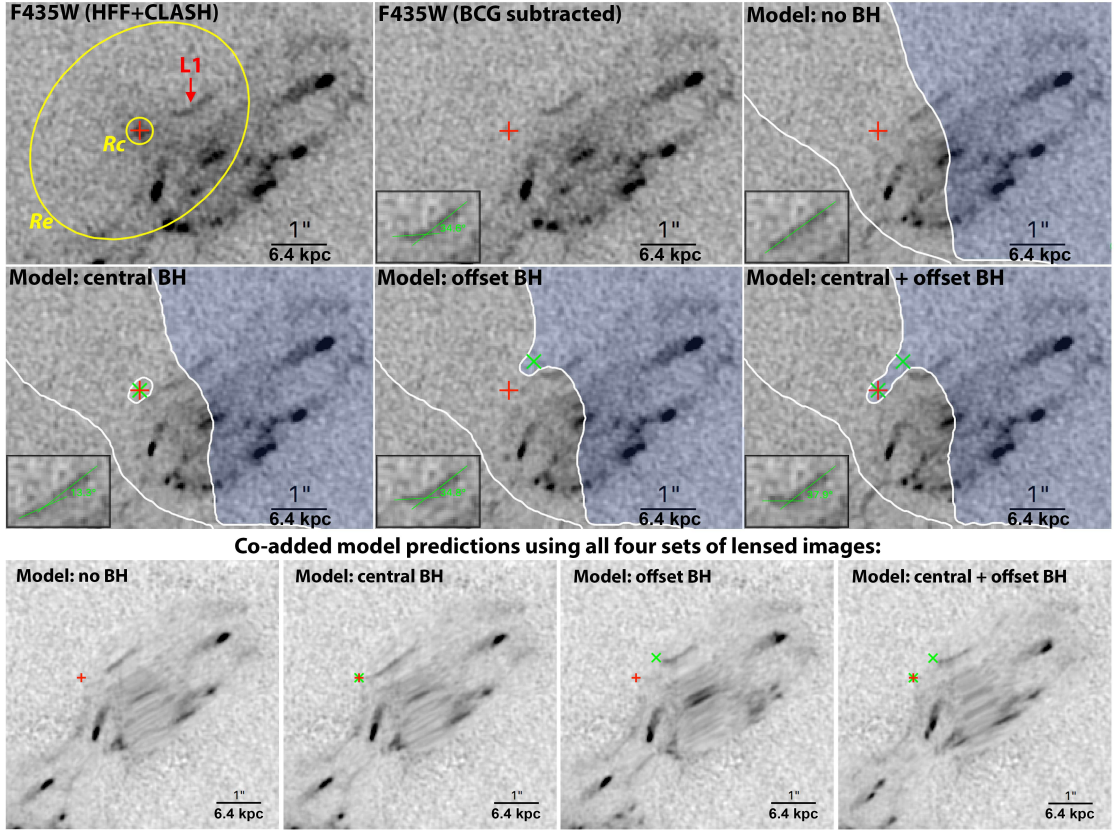}
\caption{\label{fig:result}Comparison between data and lens model predictions. {\it Upper and middle rows:} image in the F435W band that provides the best contrast between the BCG (at its faintest in this band) and the multiply-lensed HII regions of the spiral galaxy (at their brightest in this band). The projected light center of the BCG (determined in the infrared) is marked with a red cross in all the panels. The BCG's stellar light core, $R_c$, and effective radius, $R_e$, are also plotted for reference. Subtraction of the BCG light in the B-band (derived by scaling from its infrared light) is shown in the upper middle panel, enhancing the appearance of nearby lensed features, including the ``banana" shaped feature L1 in the spiral galaxy. We apply the free-form lens model to relens all the data pixels highlighted by the blue area lying to the right of the critical lensing curve (white line), so that everything to the left of this line is the model prediction. The predicted data in ``Model: no BH'' is in very good agreement with the observed data except for the curved feature L1, for which we predict a linear feature symmetrically folded over a critical curve. Adding a central black hole does not produce the observed curvature but instead lengthens the predicted image of L1. A good match is obtained by placing the black hole closer to L1, at the position indicated by the green cross, with an Einstein radius of $0\farcs15$. Adding a second black hole at the centre does not improve the fit, but is not excluded by the data. The inserts in the model-related panels indicate the deviation of the predicted image L1 from a straight line, as indicated by the two green lines. {\it Lower row:} the combined predicted images, summing over all four delensed and relensed images Sp1-4, for each of the above models.}
\end{figure*}

We start by adding a point mass \textit{at the centre} of the BCG light distribution, then delensing and relensing each of the Sp1-4 images to obtain the model prediction for L1 as well as its nearby images, similar to the process adopted in Figure \ref{fig:consistency_nobh}. As before, when relensing Sp4, we only include the part of Sp4 that is indicated by the blue-highlighted region (lower right panel of Figure \ref{fig:consistency_nobh}), and therefore only the portion of L1 to the right of the critical curve. The effect of adding this central point mass is shown in Figure \ref{fig:consistency_centralbh}. As can be seen, L1 is predicted to stretch towards the point mass and therefore to the center of the BCG, unlike that actually seen in the data. Thus, a point mass at the centre of the BCG light cannot reproduce the curvature of L1 at the location observed.

Next, we consider an offset point mass. The position and mass of this point mass have been explored as three free parameters to obtain the best fit; details of the parameter optimisation will be presented in the next section. Note that after adding a point mass either at the light centroid of the BCG or at an offset position close in projection to L1, we decrease the BCG mass by a certain amount to counteract the systematic effect of this extra point mass on lensed images close to the centre of BCG. Further away from the center, we find these modifications to the BCG mass distribution to not noticeably influence the relensed images. The relensed images from different multiple images for this case are shown in Figure \ref{fig:consistency_offsetbh}.

Finally, we investigate what constraints we can place on a point mass at the BCG light centroid in addition to the offset point mass responsible for producing the curvature of L1. As shown in Figure \ref{fig:consistency_2bh}, the presence of a central point mass with a mass equal to the offset point mass does not obviously bring about any improvement between the actual lensed images and their delensed-relensed counterparts for either L1 or other neighbouring lensed features. We conclude that the present lensing data neither favour nor exclude the possibility of having a dual point mass system. 
 
In Figure \ref{fig:consistency_centralbh}-\ref{fig:consistency_2bh}, all the bright lensed features seen in the data appear in each of the four delensed and then relensed images, and are similar with each other in each model. This good agreement clearly demonstrates that an offset point mass can plausibly reproduce the bending of L1, while a point mass centred at the BCG centroid can only elongate the image L1 while not bending it enough. A double point mass solution makes no obvious difference to the relensed images, hence no conclusion can be made concerning a second point mass at the BCG centroid. We present the different models described above together in Figure \ref{fig:result}.

\subsection{Statistical uncertainties of SMBH parameters}

\begin{figure*}[tp!]
\centering
\graphicspath{{/Users/Mandy/GoogleDrive/BH_paper/ApJ/figures/}}
\includegraphics[width=13cm]{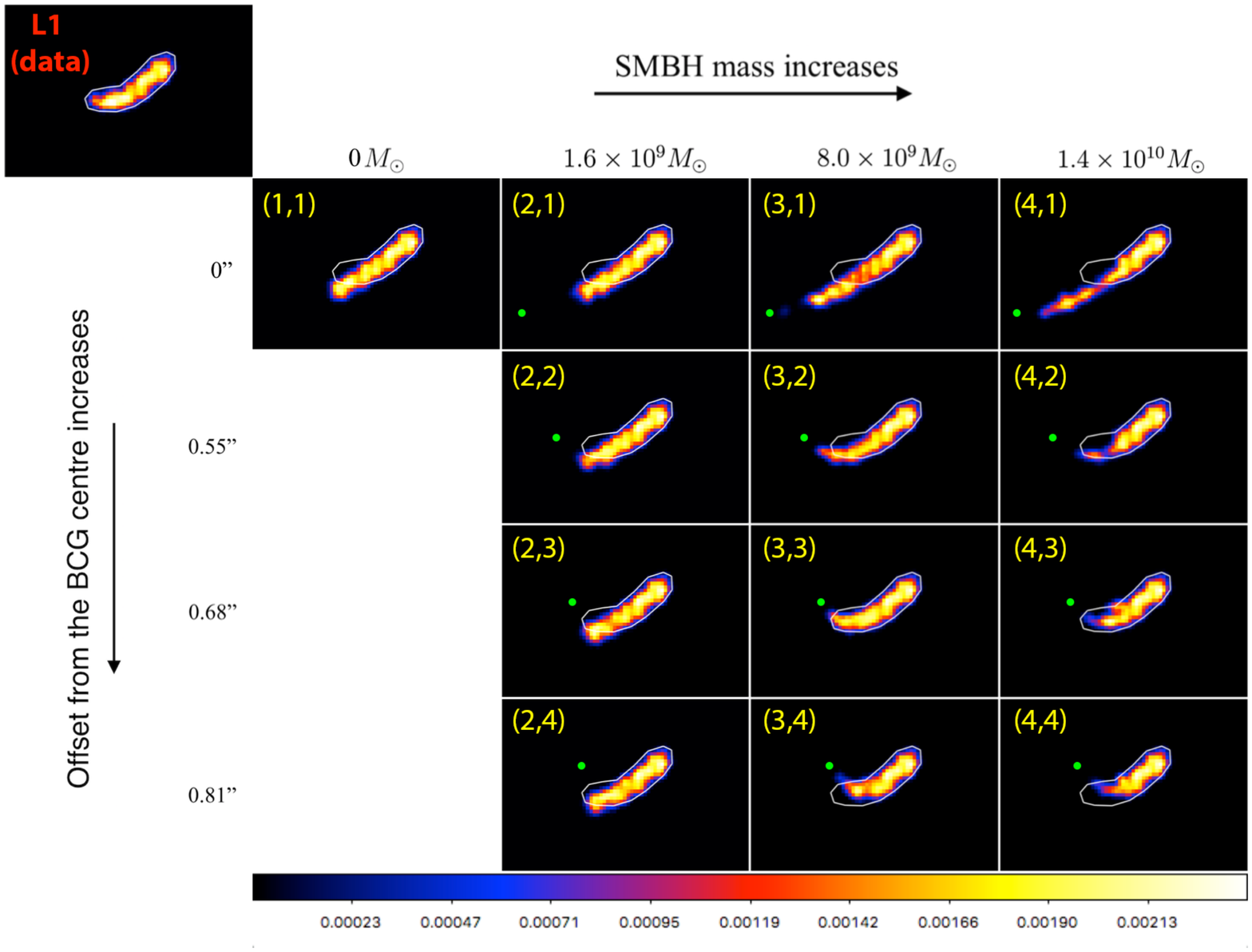}
\caption{\label{fig:manybanana}Observed versus predicted images of L1 as a function of position and mass of a supermassive black hole. Uppermost panel shows L1, enclosed within an isophotal contour that is repeated in the remaining model panels. We explore a grid of positions and masses for a point mass that when added to the WSLAP+ lens model can produce the observed curvature of L1, where the offset of the point mass from the stellar light centroid of the BCG is shown on the vertical scale and its mass shown on the horizontal scale. The position of the point mass is indicated by the green dot in each model panel. In these panels, the best match is provided by a mass of $\sim8\times10^{9}M_{\odot}$ with an offset of $\sim0\farcs7$ from the projected stellar light centre. Color bar has units of $e^{-}/s$ in the F435W band.}
\end{figure*}

\begin{figure*}[tp!]
\centering
\graphicspath{{/Users/Mandy/GoogleDrive/BH_paper/ApJ/figures/}}
\includegraphics[width=15cm]{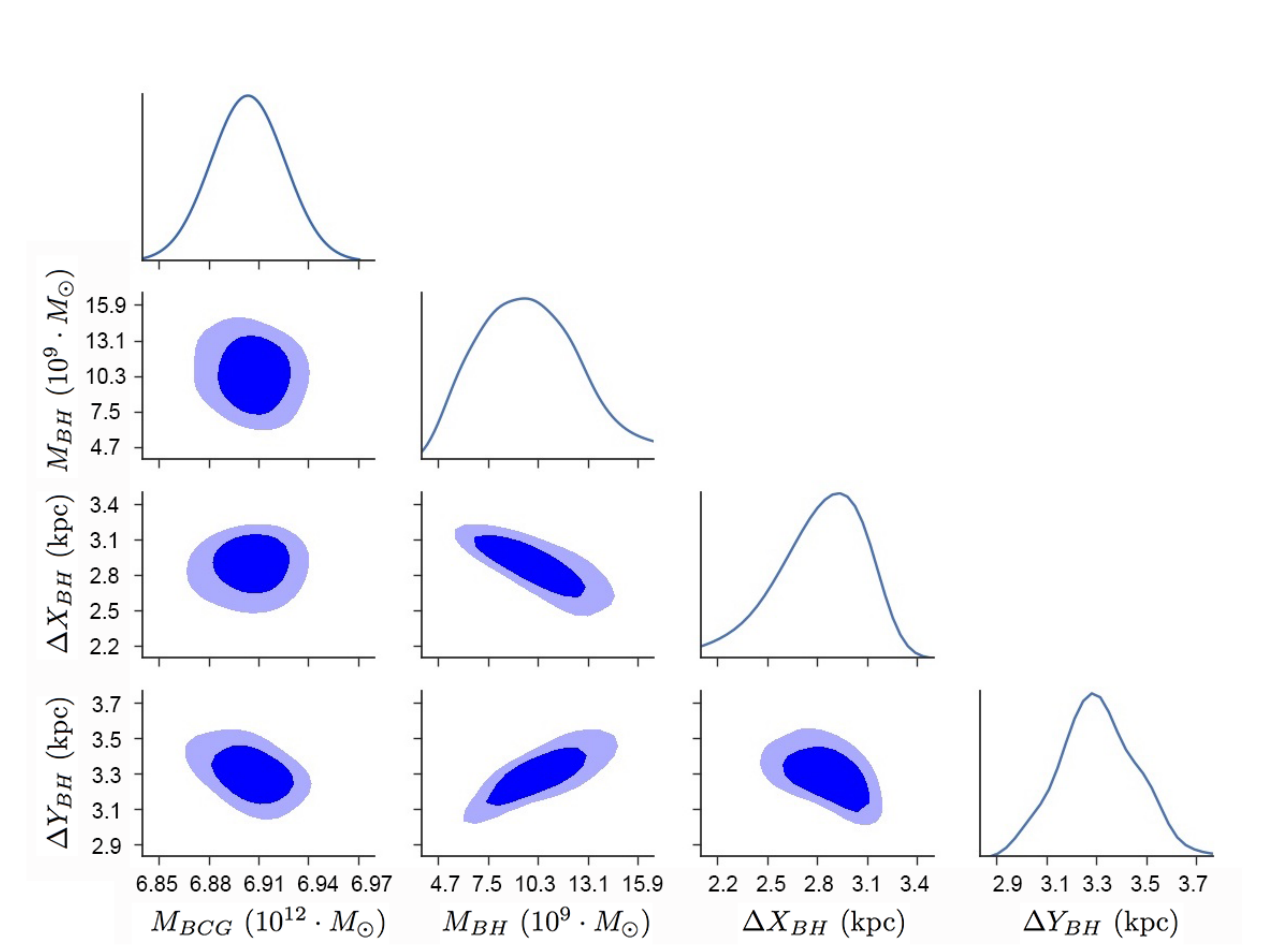}
\caption{\label{fig:Likelihood_Contours}Posterior probability distributions of BCG mass and the 3 parameters describing the black hole, its position x and y, and its mass $M_{BH}$. Contours represent 68\% and 95\% confidence levels. The BH parameters are constrained to be (within in 68\% CI) $M_{BH} = 8.4^{+4.3}_{-1.8}\times10^{9}M_\odot$ and $\Delta_{BH} = 4.4\pm0.3$ kpc.}
\end{figure*}

\begin{figure}[tp!]
\centering
\graphicspath{{/Users/Mandy/GoogleDrive/BH_paper/ApJ/figures/}}
\includegraphics[width=\linewidth]{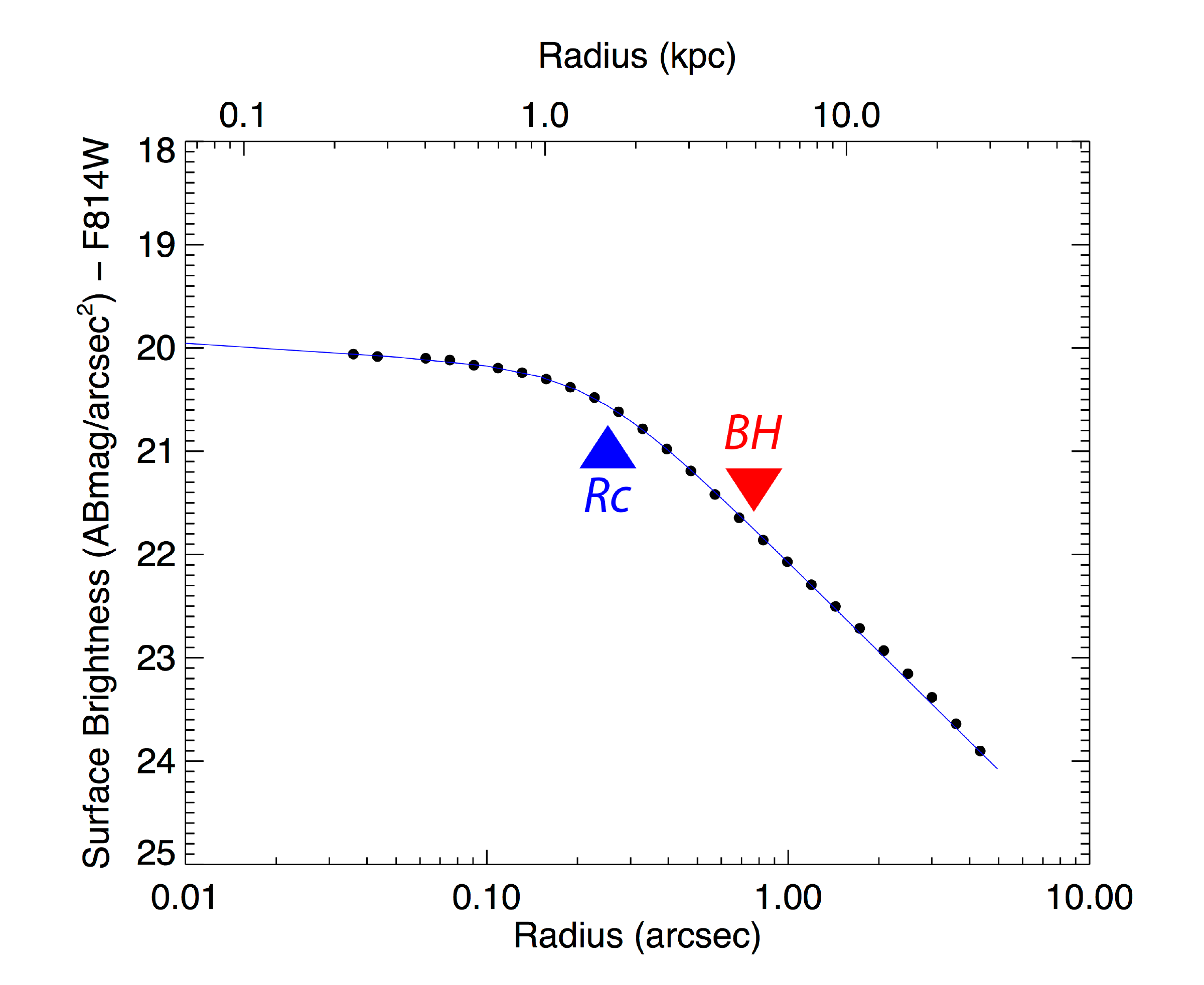}
\caption{\label{fig:Light profile with Nuker fitting line}A standard ``Nuker" profile fitted to the BCG light distribution. The profile shows a distinct break at radius of $1.5$kpc. The radial location we derive for the black hole is marked (at $\sim5$kpc), lying beyond the break.}
\end{figure}

We now describe the statistical constraints placed on the parameters of the additional point mass, which is presumably a SMBH. Figure \ref{fig:manybanana} demonstrates that the shape of the relensed L1 image is sensitive to both the position and mass of this SMBH. The models with no SMBHs predict a relatively straight L1 (position (1,1) in Fig. \ref{fig:manybanana}). Placing the SMBH at the light centre (first row with 0\arcsec offset) predicts images that are too long and straight. Therefore, we can constrain the parameters of the SMBH by comparing the actual image of L1 with its delensed and then relensed counterparts predicted by the lens model. Its position and mass ($X_{BH}$, $Y_{BH}$ and $M_{BH}$), as well as the total mass of the BCG ($M_{BCG}$), are constrained simultaneously through Markov Chain Monte Carlo (MCMC) sampling. The posterior probability distribution of the parameters $\zeta$ is sampled given the observed data $d$:
\begin{equation}
p(\zeta|d) \propto \mathcal{L}_{pos}(d|\zeta)\mathcal{L}_{L1\_angle}(d|\zeta)\mathcal{L}_{L1\_flux}(d|\zeta). 
\label{eq_posterior_prob} 
\end{equation}
$\mathcal{L}_{pos}$ is the likelihood function for the positions predicted for three bright and compact features close to the BCG centre enclosed in the green circles of Figure \ref{fig:WSLAPoutput_BCGcompare}, defined as
\begin{equation}
\mathcal{L}_{pos}(d|\zeta) \propto exp\Big[-\frac{1}{2}\sum\limits_{i=1}^3\frac{(x_{pred}^i - x_{obs}^i)^2}{0.5^2}\Big],
\label{eq_L_pos} 
\end{equation}
where we take the uncertainty of position measurements in the data as 0.5 pixel ($0\farcs015$). The position prediction of these three major features strongly constrains the total mass of the BCG. $\mathcal{L}_{L1\_angle}$ is the likelihood function for the predicted angle of L1 as defined in the manner described below. In each model with certain parameter values, we predict the shape of L1 through the delensing and relensing process as demonstrated in Figure \ref{fig:result}. According to where the critical curve passes through L1 in each model, we divide the predicted L1 into two parts, and fit a straight line to each part to obtain the angle between the two lines, $\theta_{L1pred}$. We use the same cutting point to divide L1 in the actual image, and obtain $\theta_{L1obs}$ accordingly. Then we define $\mathcal{L}_{L1\_angle}$ as
\begin{equation}
\mathcal{L}_{L1\_angle} \propto exp\Big[-\frac{1}{2}\frac{(\theta_{L1pred} - \theta_{L1obs})^2}{\sigma^2_{\theta\_L1obs}}\Big],
\label{eq_L_L1angle} 
\end{equation}
where the uncertainty $\sigma_{\theta\_L1obs}$ is propagated from the slope uncertainties of the two fitted lines to L1 data. In addition to the bending angle of L1 thus defined, we also measure the average flux density in the predicted region of L1 to make sure the surface brightness is correctly reproduced in the best-fit models. Thus we added a third likelihood function $\mathcal{L}_{L1\_flux}$ which is defined as
\begin{equation}
\mathcal{L}_{L1\_flux} \propto exp\Big[-\frac{1}{2}\frac{(\overline{f}_{pred} - \overline{f}_{obs})^2}{\sigma^2_{\overline{f}_{obs}}}\Big],
\label{eq_L_L1flux} 
\end{equation}
taking all the pixels that are relensed into the calculation.

The result of the MCMC sampling is shown in Figure \ref{fig:Likelihood_Contours}. We obtained a mass of $M_{BH} = 8.4^{+4.3}_{-1.8}\times10^{9}M_\odot$ with an offset from the BCG centre of $\Delta_{BH} = 4.4\pm0.3$ kpc. Note that this offset position is beyond the BCG stellar light core that we derive from a standard Nuker fitting using \texttt{GALFIT} \citep{Galfit2010}, for which we obtain a break radius of $1.5$ kpc as shown in Figure \ref{fig:Light profile with Nuker fitting line}. There is an observed degeneracy between the SMBH mass and its offset with respect to the BCG centre. This degeneracy is due to the fact that within a small region enclosing the optimal position for the offset SMBH, the closer the offset SMBH is to the BCG, the further away it is from L1, and hence it requires a bigger mass to cause the same degree of change on the local shear of L1.

\subsection{Adding an extended deflector}\label{extended_deflector_subsection}

\begin{figure*}[tp!]
\centering
\graphicspath{{/Users/Mandy/GoogleDrive/BH_paper/ApJ/figures/}}
\includegraphics[width=15cm]{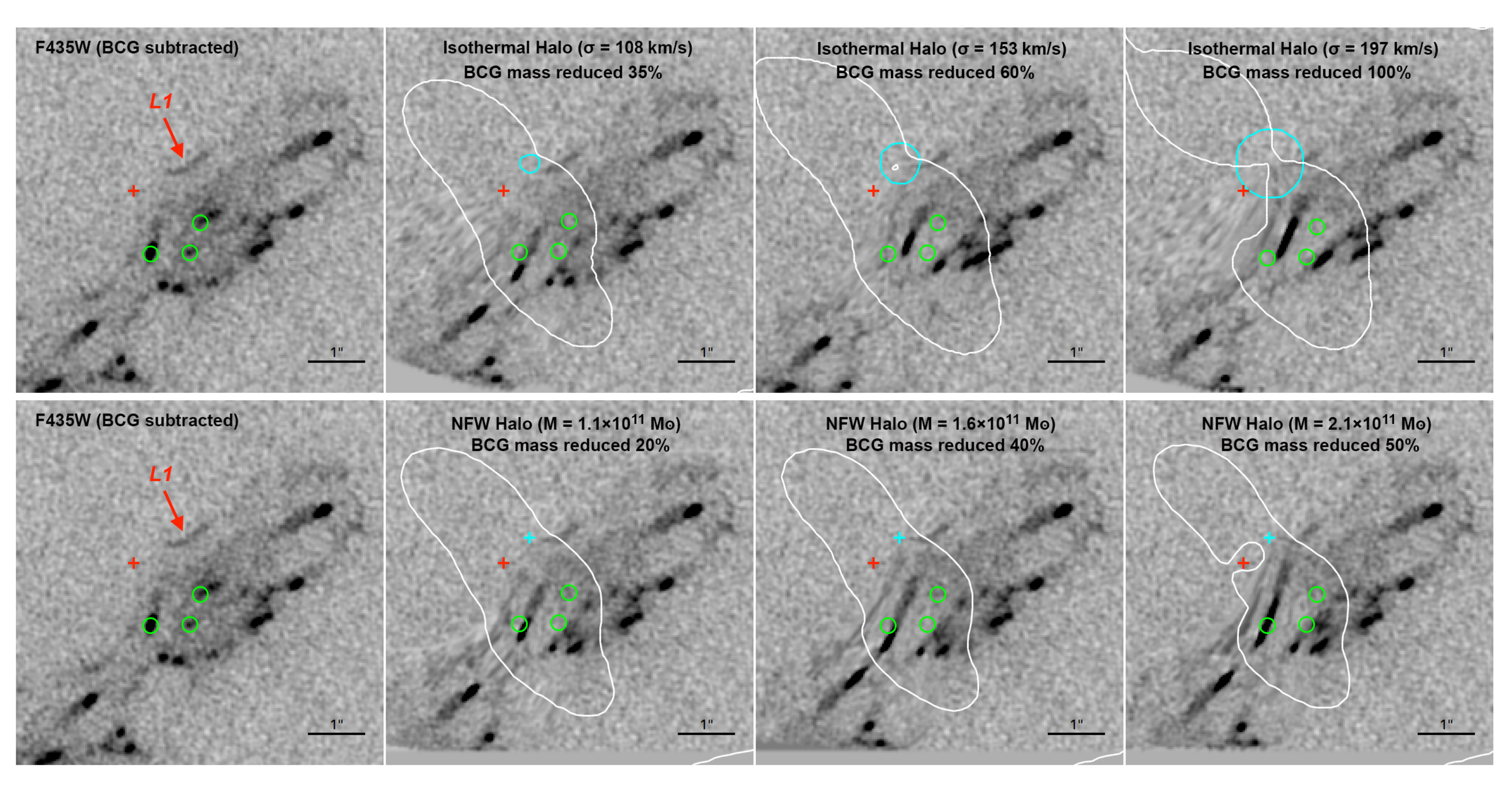}
\caption{\label{fig:isothermal_nfw_deflector}Lensing effect of an extended small substructure. Red crosses indicate the light centre of the BCG. Green circles indicate the three major lensed features in the re-lensed region, same as in Figure \ref{fig:WSLAPoutput_BCGcompare}. {\it Upper row:} replacing the offset point mass with a halo comprising a singular isothermal sphere with a velocity dispersion, and hence mass, as indicated in each panel. The cyan circles indicate the corresponding isothermal sphere's Einstein radius for a source at $z=1.4888$.  The BCG mass is scaled down as the halo mass is raised, as indicated in each panel. Although such a lens can reproduce the observed curvature of L1, its relatively extended deflection field has the effect of significantly deflecting other lensed features so that we no longer achieve good agreement between these features and the data. {\it Lower row:} as in the upper row but now but with an NFW profile representing the extended halo. Cyan crosses indicate the centres of NFW halos. Once again, although we can reproduce the observed curvature of L1, the other lensed features are perturbed so that they no longer agree with the data.}
\end{figure*}

\begin{figure*}[tp!]
\centering
\graphicspath{{/Users/Mandy/GoogleDrive/BH_paper/ApJ/figures/}}
\includegraphics[width=15cm]{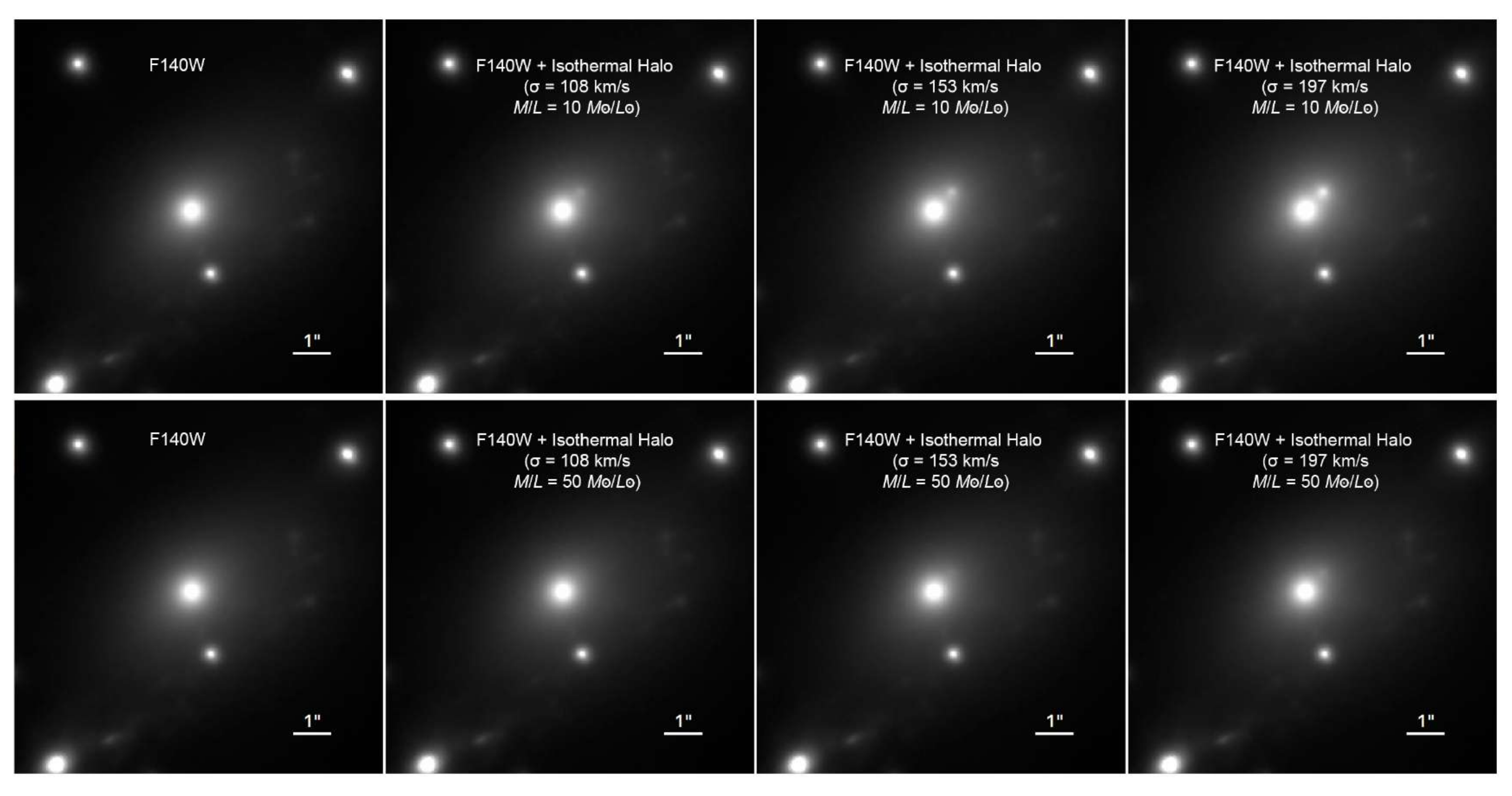}
\caption{\label{fig:isothermal_in_data}Visibility of a halo comprising a singular isothermal sphere as described in Figure \ref{fig:isothermal_nfw_deflector} for different mass-to-light (M/L) ratios. {\it Upper row:} $M/L = 10(M/L)_{\odot}$, as is the case for a typical elliptical galaxy. Three different velocity dispersions and hence masses for the halo are shown in their corresponding luminosity, with the latter computed from the bolometric correction in \cite{Buzzoni2005} as well as convolved with a PSF adopted from an isolated star in F140W band data. {\it Lower row:} same as the upper row but with a high mass-to-light ratio of $50(M/L)_{\odot}$. Despite the glare from the BCG, one should be able to see the starlight from the halo unless it has a high $M/L$ ratio above 50.}
\end{figure*}

Instead of an offset SMBH, we now examine whether a spatially extended substructure can account for the bending of L1, and if so its required parameters. For this purpose, we replace the offset SMBH with a singular isothermal halo centred at the same position. The general effect of this can be seen in Figure \ref{fig:isothermal_nfw_deflector} (upper row). Note that after adding an extended halo, the BCG's mass needs to be reduced accordingly to retain a relative satisfactory fit for all the nearby lensed features close to the BCG (also visible in Fig. \ref{fig:isothermal_nfw_deflector}). We find that although an extended halo has the effect of bending L1 in the manner desired, it also significantly influences the neighbouring details of the lensed spiral galaxy in a noticeably undesirable way. Specifically, the more massive this extended halo and therefore the greater the reduction required for the BCG mass, the worse the match between model predictions and data for the other neighbouring lenses images become. This effect can be appreciated through the relative error in model residuals obtained by (model-data)/data. In the relensed region enclosed by the critical curve shown in Figure \ref{fig:isothermal_nfw_deflector} (upper middle panel), the (model-data)/data has a standard deviation of 12.3, and a median of -0.18. In comparison, in the relensed region enclosed by the critical curve shown in Figure \ref{fig:result} (central middle panel), the (model-data)/data has a standard deviation of 7.4, and a median of -0.11. Furthermore, it can be observed that three major features indicated by green circles in Figure \ref{fig:isothermal_nfw_deflector} are poorly reproduced after an extended halo is added to the model. 

We also tried halos having different mass profiles. A NFW halo has a shallower central profile than a singular isothermal halo so that the deflection angle increases with the projected angle out to several Einstein radii, making matters even worse as shown in Figure \ref{fig:isothermal_nfw_deflector} (lower row). Later on in section \ref{lenstool_section}, we will show that a simultaneous fitting of an extended halo along with the BCG can not produce a solution more plausible than (i.e., inferior to that of) a point-mass solution as described above.

In this context, we can also ask whether the luminous stars expected to be associated with any extended deflector should be visible in contrast against the BCG light. We adopt a $M/L$ of $10(M/L)_\odot$, and convert the total luminosity $L$ to luminosity density in I band $L_I$ according to the bolometric correction estimated in \cite{Buzzoni2005}. We assume for the extended halo an elliptical galaxy having an age of 4.0 Gyr (note that the lookback time to the cluster is 5.4 Gyr), and scale the luminosity density as $L_I \simeq 0.15L$ \citep{Buzzoni2005}. Requiring the light distribution to follow the two-dimensional mass distribution of the extended deflector, we then convert $L_I$ to signals in HST F140W data with respect to the CCD inverse sensitivity (i.e. the PHOTFLAM keyword in data header). We convolve this light map with a point spread function (PSF) adopted from an isolated star (at RA = 11:49:32.697, DEC = +22:24:08.61) in the HFF F140W data. Finally, we add this converted light map to the F140W data, as shown in Figure \ref{fig:isothermal_in_data} (upper row). We expect to clearly see associated stars. Increasing $M/L$ reduces the associated starlight contrast but must reach a large value of $M/L > 50(M/L)_\odot$ to be lost in the contrast against the BCG, as shown in Figure \ref{fig:isothermal_in_data} (lower row). Although ultra-faint dwarfs can have $M/L$ as large as $\sim1000(M/L)_\odot$, these objects have very low masses (with equivalent velocity dispersion $<10$ km/s) \citep{SimonGeha2007}; we know of no objects having the required equivalent velocity dispersion of $>100$ km/s and such a large $M/L$. The arguments presented above argue against an extended halo rather than a black hole for explaining the appearance of L1.


\section{\textit{Lenstool} model of MACS 1149}\label{lenstool_section}
As an independent check of the cluster-scale lens model derived using our non-parametric, grid based method (WSLAP+), we employ the parametric lens modelling package \textit{Lenstool} \citep{lenstool} to construct an independent lens model for MACS1149. As we will show, this package also has the advantage of permitting simultaneous fitting of the BCG and a local deflector required to bend L1. We begin by allowing \textit{Lenstool} to freely describe all three main sources of deflection, namely the cluster, the BCG and the member galaxies, to find a best fit to the full set of detected multiple images, including the internal substructures of the well resolved spiral galaxy at z=1.49. Three PIEMD halos (see Eq.\ref{eq_PIEMD}), found through trial and error, are used to represent the cluster-scale mass distribution, and the BCG together with three other galaxies that are close to the lensed spiral galaxy images as pointed out in Figure \ref{fig:DelensImg} are each described by a single PIEMD halo. The rest of the cluster member galaxies are modelled by PIEMD halos scaled with respect to their luminosities as described in \cite{lenstool}. In doing so, we only restrict the BCG to be centered at its observed light centroid, so that \textit{Lenstool} is free to obtain the BCG density profile, core radius, ellipticity and position angle. The model cluster halo has the additional freedom to be centered without restriction and with any additional sub-structures according to the \textit{Lenstool} prescription. We used the same set of multiple image constraints from the WSLAP+ modelling, as listed in Table \ref{image-constraints}.  The lens model we obtained from \textit{Lenstool} is very similar to the WSLAP+ model that we use for all the previous analyses, boosting confidence in our results as presented above. 

\begin{figure*}[tp!]
\centering
\graphicspath{{/Users/Mandy/GoogleDrive/BH_paper/ApJ/figures/}}
\includegraphics[width=15cm]{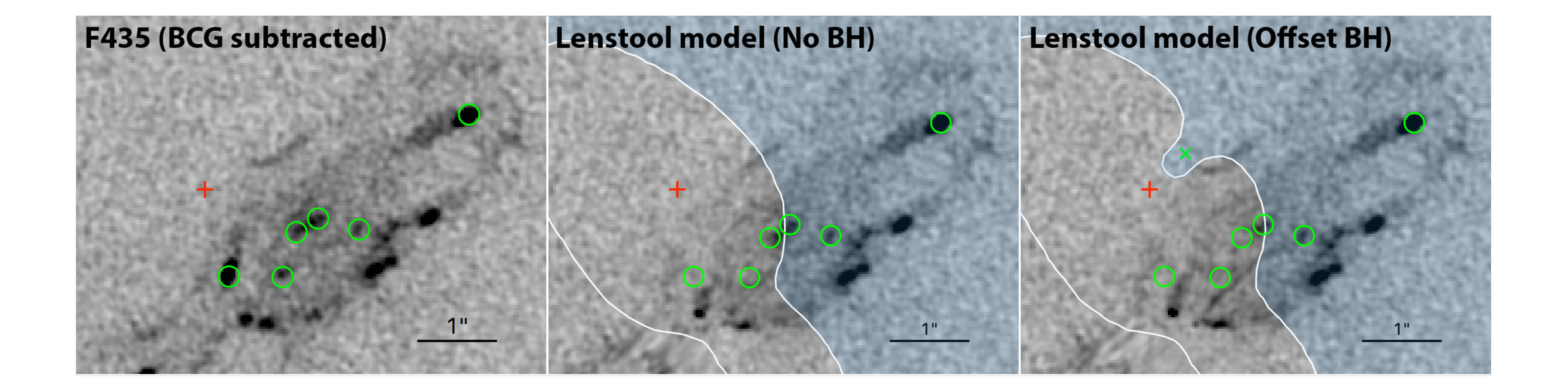}
\caption{\label{fig:lenstool_delensrelens}Relensed predictions from \textit{Lenstool} model. Green circles point out three major features close to the centre of the BCG, and they are plotted in both model panels to indicate the offset of the model predicted positions. Without adding an offset dark point mass, the parametric package \textit{Lenstool} produces a consistent model with WLSAP+. A critical curve passes through L1, leaving the prediction from the right half of L1 very straight (middle panel). We later add a point mass in the lens modelling with its position and mass being a free parameter, and re-run the model construction with leaving all parameters of the BCG free while holding the other parameters fixed. \textit{Lenstool} produces a similar result as WSLAP+ with an extra point mass, with offset from the BCG of $0\farcs66\pm0.11$ and a higher best fit mass of $1.2^{+0.19}_{-0.22}\times10^{10}M_\odot$ (relensed image shown in the right panel).}
\end{figure*}

The best fit model from \textit{Lenstool} has a mean rms dispersion between the predicted and observed centroids for each set of multiply-lensed images (used to constrain the lens model) as measured in the image plane of $<rms_\textit{i}>=0\farcs25$. Compared with WSLAP+ model, this \textit{Lenstool} model has a higher large-scale accuracy towards the outer region of the cluster. We show this \textit{Lenstool} solution in Figure \ref{fig:DelensImg} (4th row) and also Figure \ref{fig:lenstool_delensrelens}. The corresponding de-lensed source plane images of Sp1-4 from \textit{Lenstool} (Figure \ref{fig:DelensImg}, 4th row) show a very high degree of agreement in shape and position, exceeding the accuracy of our WSLAP+ solution. Most importantly for our conclusions regarding the curved image, L1, the \textit{Lenstool} critical curves around the BCG can be seen to be very similar to our previous WSLAP+ solution, passing through the center of the curved L1 image. The similar critical curves at the position of L1 found using either WSLAP+ or \textit{Lenstool} indicates that our conclusion does not depend strongly on the assumed mass profile of the BCG. As shown in Figure \ref{fig:lenstool_delensrelens}, the relensed image of L1 is quite straight just like our previous solution, being highly sheared because it straddles the critical curve as discussed in detail in section \ref{shear_section}. This critical curve also passes through the center of the other nearby highly sheared images and close pairs that evidently straddle this critical curve, which are pointed out by the encircled features in Figure \ref{fig:lenstool_delensrelens}. These pairs provide a convenient measure of the accuracy of the lens model in the region around the BCG. Without an addition of a local deflector to bend L1, the mean rms dispersion between the predicted and observed centroids for each set of these multiply-lensed images (i.e., the circles in Figure \ref{fig:lenstool_delensrelens}) as measured in the image plane is $<rms_\textit{i}>=0\farcs2$. 

\subsection{Adding a local point mass}
\begin{figure*}[tp!]
\centering
\graphicspath{{/Users/Mandy/GoogleDrive/BH_paper/ApJ/figures/}}
\includegraphics[width=15cm]{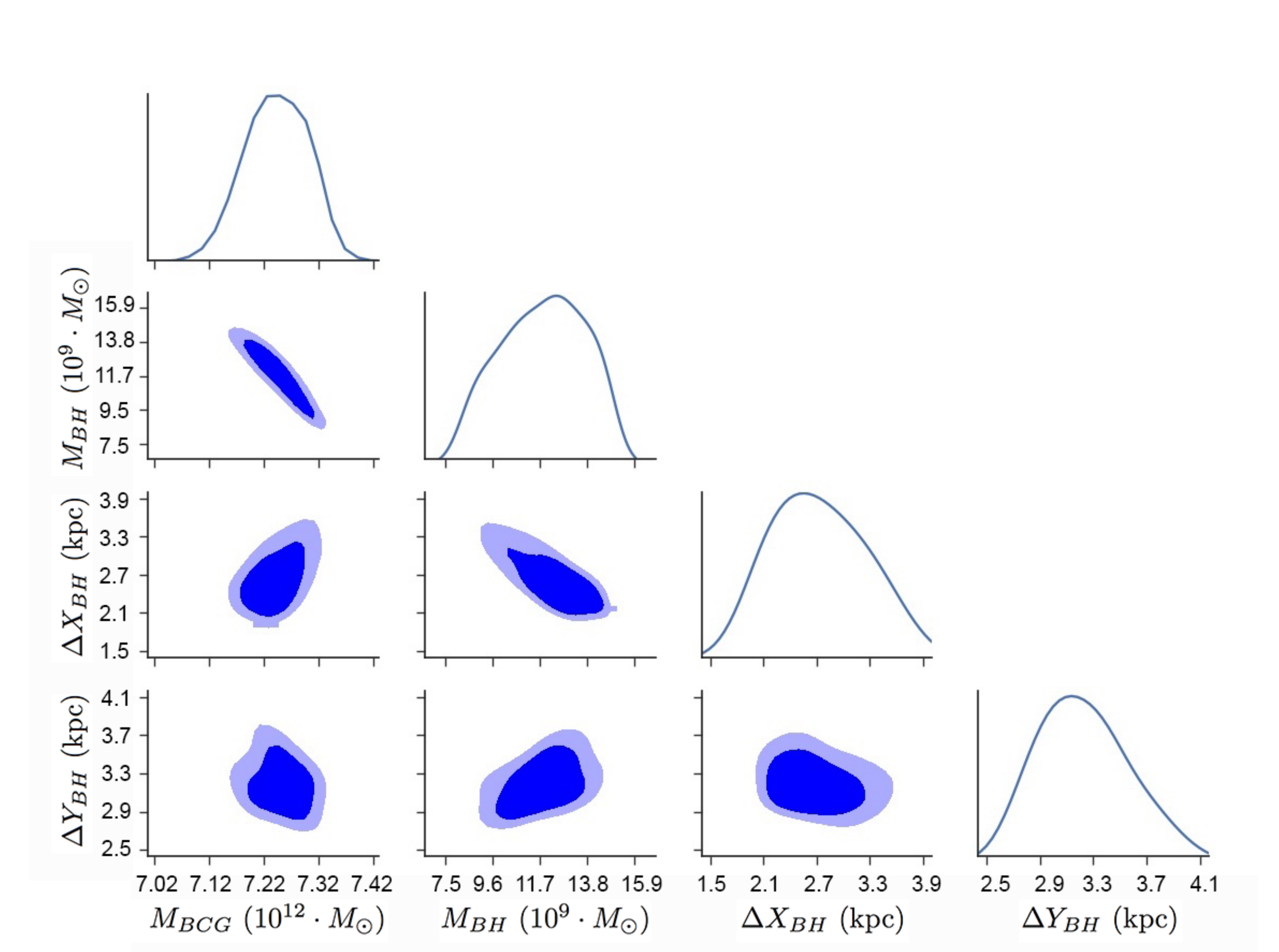}
\caption{\label{fig:lenstool_mcmc}Posterior probability distributions of BCG mass and the 3 parameters describing the black hole, its position x and y, and its mass $M_{BH}$ from \textit{Lenstool} model. Similar to Figure \ref{fig:Likelihood_Contours}, contours represent 68\% and 95\% confidence levels. In this \textit{Lenstool} solution, the BH parameters are constrained to be (within in 68\% CI) $M_{BH} = 1.25^{+0.14}_{-0.27}\times10^{10}M_\odot$ and $\Delta_{BH} = 4.2^{+0.7}_{-0.5}$kpc.}
\end{figure*}
Next, we require \textit{Lenstool} to find a best fit solution with a point mass added to produce the observed curvature of L1. As the goal here is to reproduce L1 based on a good cluster lens model we have already obtained, we restrict the model constraints to the images in the vicinity of the BCG, including L1, while excluding images from different background galaxies and images far away from the BCG centre. We in turn fix the three cluster PIEMD halos as they are not well constrained by the restricted set of image constraints. The BCG's PIEMD halo position is fixed to be at its observed light centroid with all other parameters free to vary. We also fix the PIEMD parameters for the three other galaxy halos except for their velocity dispersions (i.e. total mass). The point mass added is allowed to lie within a $1\arcsec\times1\arcsec$ square centered on the centroid of the BCG, with a wide uniform prior on its mass, relevant for SMBHs. We use ``forme = -3'' in the \textit{Lenstool} parameter file so that the minor and major axes of L1 is included as constraints. The optimisation of the aforementioned free parameters is performed in the image plane. 

The best fit \textit{Lenstool} solution found this way is centered near the radius of curvature of L1, with a mass whose Einstein radius is in correspondence with the approximate curvature radius of L1. The effect on the shape of the critical curve is the same as what we obtained previously with WSLAP+, forming an indentation near L1 as shown in Figure \ref{fig:lenstool_delensrelens}. \textit{Lenstool} constrains this point mass offset from the BCG as $\Delta_{BH} = 4.2^{+0.7}_{-0.5}\textrm{kpc}$ with a higher best fit mass of $M_{BH} = 1.25^{+0.14}_{-0.27}\times10^{10}M_\odot$. These parameters are in good agreement with those derived in section \ref{result_section} based on WSLAP+ lens model within the uncertainties. The posterior probability distributions of the SMBH parameters and BCG mass obtained from \textit{Lenstool} model are shown in Figure \ref{fig:lenstool_mcmc}. The mean rms dispersion between the predicted and observed centroids for each set of the multiply-lensed images circled in Figure \ref{fig:lenstool_delensrelens} as measured in the image plane remains unchanged at $<rms_\textit{i}>=0\farcs2$, indicating that the addition of a point mass has not appreciably perturbed the neighbouring lensed images.

\subsection{Adding a local extended halo}
\textit{Lenstool} can also readily illuminate the question of an extended deflector, allowing us to obtain constraints on its mass, profile slope, core and truncation radii. We add an extended PIEMD halo to the \textit{Lenstool} input (without a point mass), and first place this extended galaxy within the lens plane (z = 0.543). The best fit result requires the extra halo to be cuspy and highly truncated with a core radius of $0.8$ kpc and truncation radius of $3.7$ kpc. This best-fit halo has a mass of $2.3\times10^{10}M_\odot$, producing a high surface mass density giving its small truncation radius. In this case, the mean rms dispersion between the predicted and observed centroids for each set of the multiply-lensed images circled in Figure \ref{fig:lenstool_delensrelens} as measured in the image plane is $<rms_\textit{i}>=1\farcs3$, much poorer than in the case of adding a point mass. Thus, although a massive and sharply truncated halo can bend L1 to the degree required, it also perturbs neighbouring lensed images in an undesirable manner, as we found earlier by adding a halo in the WSLAP+ model. With the Bayesian likelihood values output by \textit{Lenstool}, we also compute the Bayesian Information Criterion (BIC, ~\citealt{BIC1978}) for three models: (1) model without an extra deflector local to L1, (2) model with an extended halo as a local deflector, and (3) model with a point mass as a local deflector. Their BIC values are 17.97, 27.25, and -0.80, respectively. The model with a point mass is preferred as it has the lowest BIC value, while the model with an extended halo has the highest BIC value and is disfavoured.

\begin{figure*}[tp!]
\centering
\graphicspath{{/Users/Mandy/GoogleDrive/BH_paper/ApJ/figures/}}
\includegraphics[width=12.5cm]{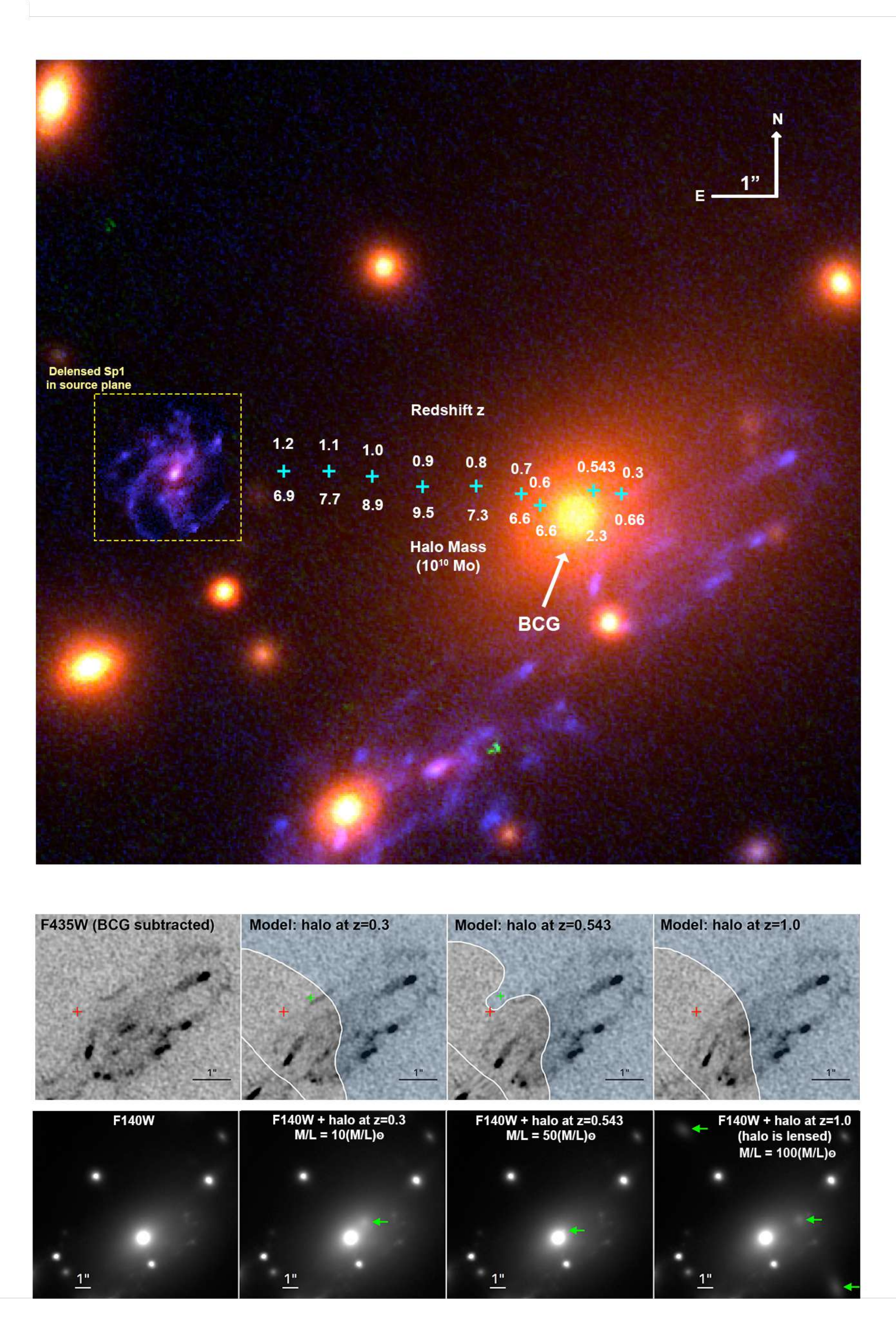}
\caption{\label{fig:gal_different_z}Adding an extended PIEMD halo at different redshifts. {\it Upper row:} best fit position and mass of an extended halo at different redshifts. The delensed image of Sp1 in the source plane is overplotted on this color data image, and the cyan crosses indicate the best fit position of the extended halos with their redshifts marked on the top and best fit mass marked on the bottom. Those extended halos align in the region between the projected source position and the position of L1, with generally bigger masses towards higher redshifts while moving closer to the source position. {\it Middle row:} relensed predictions with an extended halo at redshift 0.3, 0.543 and 1.0. The red cross is the position of the BCG's light centre and the green cross is the position of the extended halo. Note that in the case of $z=1.0$, the extended halo is located outside the field of view showing in this model prediction region. {\it Lower row:} halo starlight added to F140W data. The extended halo with mass shown in panel a is converted to starlight (with $M/L = 10(M/L)_\odot$, $M/L = 50(M/L)_\odot$, and $M/L = 100(M/L)_\odot$ respectively) by the method described in Chapter 6, similar to Figure \ref{fig:isothermal_in_data}. Note that when the halo is at z = 1.0, it is strongly lensed by the cluster and forms three multiple images.}
\end{figure*}

We then explore the addition of an extended deflector at a different redshift ranging over the redshift interval 0.3 to 1.2. The best fit mass and position for this extra halo at different redshifts are shown in Figure \ref{fig:gal_different_z}. We conduct this analysis by manually adding a second lens plane containing only the extended deflector to the \textit{Lenstool} modelling, and it is not a full multi-plane analysis, which is currently not supported by \textit{Lenstool}. As can be seen, this halo is required to lie along the line of sight between the projected source position and the position of L1, increasing in mass and moving closer to the source position with increasing redshift. All the solutions prefer a small truncation radius ($\sim3$ kpc) comparable to the radius of curvature of the lensed image, with an unconstrainably small core radius. As can be seen, as the halo moves significantly away from the cluster redshift (either towards lower or higher redshifts), it cannot alter the shape of the critical curve in the cluster lens plane sufficiently so as to produce the observed curvature of L1. 

Similar to Figure \ref{fig:isothermal_in_data}, we also explore the visibility of this extended mass in contrast to the BCG light. We convert the mass distribution to light distribution with a $M/L = 10(M/L)_\odot$, $M/L = 50(M/L)_\odot$, and $M/L = 100(M/L)_\odot$ for the three halos at redshift 0.3, 0.543 and 1.0, respectively, following the method described in section \ref{extended_deflector_subsection}. Note that when the extended halo is located beyond the cluster redshift of 0.543, the halo image is lensed by the cluster which will impose a magnification of $\mu>2$. In the case of a halo at z=1.0, this halo is lensed into three multiple images with a total magnification of $\mu\simeq8.5$. We add the converted light maps for the extended halo at z=0.3, z=0.543, and z=1.0 to F140W data, as shown in Figure \ref{fig:gal_different_z} (lower row). If at z=0.3 and having $M/L = 10(M/L)_\odot$, this halo should be clearly visible in the data; a much larger $M/L$ is required to make any such halo undetectable. For halos at $z>0.543$, a large $M/L$ of $>100(M/L)_\odot$ is required for the extended halo to be invisible in the data. In combination with its even more unfavourable ability to bend L1 to the degree required, both these factors make an intervening extended halo untenable.


\section{Discussion}\label{discussion}
\subsection{Rocket effect}
The simplest explanation for a dark point mass necessary to explain L1 is a SMBH hosted by the BCG. The central light profile of the BCG galaxy has a pronounced flattening within a few kpc of the light center, similar to other well known BCG galaxies where a stellar ``core" is claimed \citep{Postman2012_largecore}. A convincing explanation for such cores involves prolonged binary SMBH merging so that stars are scattered away, particularly stars on radial orbits, an idea supported by stellar dynamical data for the cluster NGC 1399 \citep{Bender2007}. The BCG stellar core radius we derive of $1.5$ kpc (see Figure \ref{fig:Light profile with Nuker fitting line}) is in line with other similarly luminous elliptical galaxies. 

The location of the compact mass beyond the stellar core may imply ejection of the SMBH by the ``rocket effect" \citep{Bekenstein.RocketEffect.1973,Begelman.Blandford.Rees.Nature1980}, where a preceding binary phase that may be responsible for the flattened core has resulted in the coalescence and ejection of the resulting merged black hole. In the calculations to follow, we estimate the initial kick velocity after the coalescence of the black hole binary and the time elapsed since the kick. 

We first assume the mass associated with the BCG is spherically symmetrical. We take the two-dimensional BCG mass distribution, and de-project  it using the inverse Abel transform to its three-dimensional mass distribution:
\begin{equation}
\rho(r) = \frac{1}{\pi}\int_{R_{max}}^{r}\frac{1}{\sqrt{R^2-r^2}}\frac{dI(R)}{dR}dR, 
\label{eq_inverseAbel} 
\end{equation}
where $\rho(r)$ is the de-projected three-dimensional mass density and $I(R)$ is the two-dimensional surface mass density obtained from our lens model. We assume that the offset SMBH is at rest at the best-fit position of our model ($\sim4.4 \textrm{kpc}$ from the centroid of the BCG), and that the distance between the SMBH and the centroid of the BCG is the projected distance observed (i.e., the minimal true distance). We then calculate the minimum initial kick velocity by considering that the SMBH is only slowed down by gravitational force due to the gravitational potential of the BCG. Specifically, we obtain the velocity of the SMBH at any given radial position r by
\begin{equation}
\frac{1}{2}v(r)^2 = 4\pi G \Big(\frac{1}{D}\int_{0}^{D}\rho(r\prime) r\prime^2dr\prime - \frac{1}{r}\int_{0}^{r}\rho(r\prime) r\prime^2dr\prime\Big), \label{eq_kickvelocity}
\end{equation}
where $D=4.4$ kpc.  The right-hand side of eq. \ref{eq_kickvelocity} is the difference between the gravitational potential energy at radius r and radius $D=4.4$ kpc. When r approaches 0, we obtain the initial kick velocity by \begin{equation}
\frac{1}{2}v_{kick}^2 = \frac{4\pi G}{D}\int_{0}^{D}\rho(r\prime) r\prime^2dr\prime. 
\end{equation}
We calculate the time elapsed since the kick using 
\begin{equation}
t = -\int_{0}^{D}\frac{1}{a(r)}\frac{dv(r)}{dr}dr, \label{eq_kicktime} 
\end{equation}
where $v(r)$ is obtained from eq. \ref{eq_kickvelocity} and the gravitational de-acceleration $a(r)$ is obtained by
\begin{equation}
a(r) = \frac{GM(r)}{r^2} = \frac{4\pi G}{r^2}\int_{0}^{r}\rho(r\prime) r\prime^2dr\prime. 
\label{eq_acceleration} 
\end{equation}

We find $v_{kick} = 314$ km/s and $t = 2.0\times10^7$ years. Theoretical studies suggest a kick velocity of a few hundred km/s \citep{kick_estimate}, demonstrating that the ``rocket effect'' is a viable explanation for the offset of the SMBH from the centre of its host galaxy. Empirically, several cases of spatially offset SMBH are now known in the SDSS survey \citep{Lena2014,Kim2016,Chiaberge2016}, linked to host galaxy merging, with a median offset of $4.6$ kpc \citep{Comerford2016} that is similar to our measured offset.

\subsection{Implication for SMBH-galaxy co-evolution}
The BCG is a very luminous galaxy with a V-band magnitude of $M_V = -24.1$ and an integrated luminosity of $3.9\times10^{11}L_{V\odot}$ within an aperture of radius 30 kpc. Assuming a standard initial mass function viewed at a look back time of $5$ Gyrs \citep{Chabrier2003}, we obtain a stellar mass of $6.3\times10^{11}M_\odot$. By comparison, in the WSLAP+ lens model, the total projected mass within a cylinder of $r<30$ kpc centred on the BCG is $7.1\times 10^{12}M_\odot$, contributed by both the BCG and other cluster matter along the line of sight. The SMBH mass of $M_{BH} = 8.4^{+4.3}_{-1.8}\times10^{9}M_\odot$ is in agreement with the local $M_{BH}-L_V$ relation between the SMBH mass and host galaxy V-band luminosity \citep{Lauer2007}. 

This SMBH mass measurement seems to show that the $M_{BH}-L_V$ relation extends to an epoch when the Universe was half of its present age. Therefore, at face value, our work provides support for the co-evolution between SMBHs and their host galaxies. The BCG studied in this paper is a giant elliptical galaxy residing at the center of a galaxy cluster; it may not grow much more in mass over the next 5 Gyrs. That this object follows the $M_{BH}-L_V$ relation may not therefore provide a strong argument that SMBHs necessarily grow in step with their host galaxies over cosmic times; i.e., it may already closely resemble BCGs in the local universe. To shed light on the co-evolution of SMBHs and their host galaxies, what is needed therefore are measurements of SMBH masses for distant galaxies that will still grow substantially in mass to the current epoch.


\section{Conclusion}\label{conclusion}
Direct and reliable determinations of the masses of SMBHs are restricted to the low-redshift Universe ($z<0.3$), where stellar or gas kinematics in the close vicinity of the SMBH can be spatially resolved \citep{Ferrarese&Ford1999,Kormendy2004}. At larger distances, we have to resort to reverberation mapping; the veracity of the masses thus inferred for SMBHs remains poorly understood. For galaxies where both these technique are not possible, we have to resort to scaling relations, the veracity of which is even more questionable \citep{Graham2016}. Nevertheless, based on these measurements, the masses of local SMBHs are found to correlate with the masses of the bulge components of their host galaxies, indicating a co-evolution history of SMBHs and their host galaxies \citep{Kormendy_Ho2013}. Bright QSOs at high redshifts, however, are inferred to harbour extremely massive SMBHs, suggesting a rapid early growth of SMBH in contradiction with the idea of a co-evolution with their host galaxies \citep{Wu2015,BennyScience}. To understand the growth of SMBH masses over cosmic time, what is clearly needed is the ability to accurately measure SMBH masses over cosmic history.

Gravitational lensing is a promising method to directly measure SMBH masses beyond the local Universe. In the case of lensing by a single foreground galaxy with near-perfect alignment between the source and the lensing galaxy, a central de-magnified image is generic to lensing, generated within the Einstein radius of the lensing galaxy, such that the larger the SMBH mass the more this central image is attracted towards the SMBH and de-magnified \citep{mao2001,Rusin2005,Hezaveh2015}. In such cases, however, there is an inherent degeneracy between the mass of a SMBH and the central mass profile of its host galaxy. In situations where a background galaxy is lensed by a foreground galaxy cluster such that an individual lensed feature appears close to the center of a cluster member, as in the case presented in this paper, an unambiguous case can be made for a SMBH and its parameters can be directly determined.

Specifically, based on a ``banana-shaped'' lensed feature L1 that is located close to the centre of the BCG in MACS 1149, we constrain the position and mass of an offset SMBH hosted by this BCG. L1 is the closest image to the centre of the BCG, and exhibits a radius of curvature of only $0\farcs6$. L1 is consistently predicted to have a straight appearance owing to the large shear at its position. To bend L1 on the small scale as observed in the data, a local deflector is required in addition to the combined cluster and member galaxy lens model to alter the shape of the critical curve cutting through L1. By adding a point mass of $M_{BH} = 8.4^{+4.3}_{-1.8}\times10^{9}M_\odot$ offset by $\sim4.4$ kpc from the centre of the BCG, we successfully reproduce the curvature of L1. A highly truncated extended halo (i.e., a galaxy) with a velocity dispersion $>100$ km/s can also bend L1 to the degree required. Such a galaxy, however, deflects neighbouring lensed images from the same background galaxy in an undesirable way. Furthermore, the halo is required to have a large mass-to-light ratio ($>50(M/L)_\odot$) to be invisible in the deep HFF data; no galaxy having such a large mass along with such a high mass-to-light ratio is known. The point-mass solution can be interpreted as an offset SMBH ejected from the BCG centre due to the asymmetric gravitational wave radiation in a merger event (the ``rocket effect''). To bring the SMBH to its present position, the kick should have occurred at least $2.0\times10^7$ years ago, and have had an initial velocity of $>314$ km/s. These estimations are in agreement with theoretical predictions for typical ``rocket effect'' events.

\acknowledgements
The authors would like to thank the anonymous referee for providing detailed and constructive suggestions to improve this paper. J.L. acknowledges support from the Research Grants Council of Hong Kong through grant 17319316. J.L. also acknowledges a Seed Fund for Basic Research from the University of Hong Kong. T.B. was supported by a Visiting Research Professor Scheme from the University of Hong Kong, during which major parts of this work were conducted. J.M.D. acknowledges the support of projects AYA2015-64508-P (MINECO/FEDER, UE), AYA2012-39475-C02-01, and the consolider project CSD2010-00064 funded by the Ministerio de Economia y Competitividad. Y.O. is supported by the Ministry of Science and Technology (MOST) of Taiwan, MOST 106-2112-M-001-008-.

\clearpage
\appendix
\renewcommand*{\thetable}{\Alph{section}\arabic{table}}
\renewcommand*{\thefigure}{\Alph{section}\arabic{figure}}

\section{Information of multiply-lensed images}
\setcounter{figure}{0}
\setcounter{table}{0}
We plot below the locations of multiply-lensed images used as constraints for lens modelling, and we tabulate their coordinates.

\begin{figure*}[tp!]
\centering
\graphicspath{{/Users/Mandy/GoogleDrive/BH_paper/ApJ/figures/}}
\includegraphics[width=12.5cm]{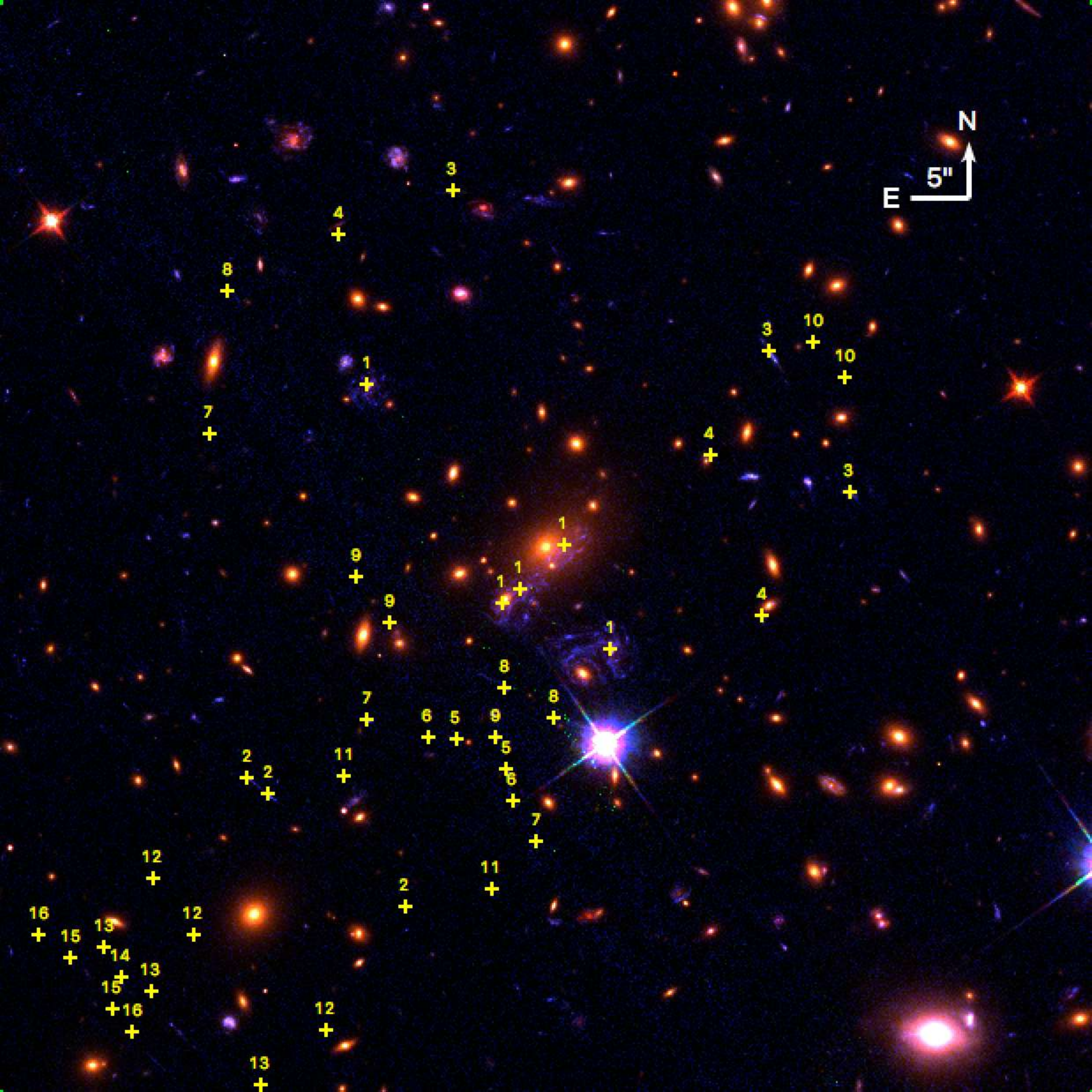}
\caption{\label{fig:image_ID} Locations of the 16 multiply-lensed image systems used as constraints for constructing both WSLAP+ and \textit{Lenstool} models. The image shown here is with the WSLAP+ modelling FOV of $1.\arcmin6\times1.\arcmin6$. System 1 corresponds to the background spiral galaxy at $z=1.4888$. We used more than 20 individual features from system 1 as constraints, as listed in Table \ref{image-constraints}.}
\end{figure*}

\startlongtable
\begin{deluxetable*}{c   c   c}
\tablecaption{Coordinates of multiply-lensed images used as constraints for lens modelling. Locations of the multiply-lensed image systems are shown in Figure \ref{fig:image_ID}. \label{image-constraints}}
\tablehead{
\colhead{Image ID} & \colhead{RA} & \colhead{DEC}}
\startdata
1.1.1 & 11:49:35.283 & +22:23:45.64 \\ 
1.1.2 & 11:49:35.213 & +22:23:43.35 \\ 
1.1.3 & 11:49:35.574 & +22:23:44.27 \\ 
1.1.3 & 11:49:35.453 & +22:23:44.82 \\ 
1.1.3 & 11:49:35.370 & +22:23:43.94 \\ 
1.1.3 & 11:49:35.474 & +22:23:42.68 \\ 
1.1.4 & 11:49:35.158 & +22:23:44.16 \\ 
1.1.5 & 11:49:35.558 & +22:23:46.86 \\ 
1.1.6 & 11:49:35.384 & +22:23:47.09 \\ 
1.1.7 & 11:49:35.307 & +22:23:48.19 \\ 
1.1.8 & 11:49:35.187 & +22:23:46.70 \\ 
1.1.9 & 11:49:35.414 & +22:23:45.99 \\ 
1.1.10 & 11:49:35.479 & +22:23:47.63 \\ 
1.1.11 & 11:49:35.639 & +22:23:45.96 \\ 
1.1.12 & 11:49:35.144 & +22:23:46.50 \\ 
1.1.13 & 11:49:35.349 & +22:23:46.37 \\ 
1.1.14 & 11:49:35.319 & +22:23:42.76 \\ 
1.1.15 & 11:49:35.250 & +22:23:46.37 \\ 
1.1.16 & 11:49:35.298 & +22:23:44.58 \\ 
1.1.17 & 11:49:35.254 & +22:23:44.74 \\ 
1.1.18 & 11:49:35.265 & +22:23:43.42 \\ 
1.1.19 & 11:49:35.272 & +22:23:47.90 \\ 
1.1.20 & 11:49:35.323 & +22:23:47.42 \\ 
1.1.21 & 11:49:35.177 & +22:23:45.36 \\
1.1.22 & 11:49:35.111 & +22:23:44.51 \\ 
1.1.23 & 11:49:35.214 & +22:23:47.66 \\ 
1.1.24 & 11:49:35.120 & +22:23:45.71 \\ 
1.1.25 & 11:49:35.498 & +22:23:45.73 \\
1.1.26 & 11:49:35.541 & +22:23:45.11 \\
1.1.27 & 11:49:35.277 & +22:23:47.21 \\   
1.2.1 & 11:49:35.858 & +22:23:50.81 \\ 
1.2.2 & 11:49:35.942 & +22:23:51.02 \\ 
1.5.3 & 11:49:36.026 & +22:23:48.10 \\ 
1.2.4 & 11:49:35.883 & +22:23:51.95 \\ 
1.2.5 & 11:49:35.824 & +22:23:48.62 \\ 
1.2.6 & 11:49:35.798 & +22:23:49.99 \\ 
1.2.7 & 11:49:35.705 & +22:23:51.48 \\ 
1.2.8 & 11:49:35.754 & +22:23:52.35 \\ 
1.2.9 & 11:49:35.896 & +22:23:49.78 \\ 
1.2.11 & 11:49:35.872 & +22:23:47.73 \\ 
1.2.12 & 11:49:35.754 & +22:23:52.92 \\  
1.2.13 & 11:49:35.840 & +22:23:50.26 \\ 
1.5.14 & 11:49:36.076 & +22:23:48.93 \\ 
1.2.15 & 11:49:35.801 & +22:23:51.35 \\ 
1.2.16 & 11:49:35.913 & +22:23:50.50 \\ 
1.2.17 & 11:49:35.889 & +22:23:50.94 \\ 
1.2.19 & 11:49:35.718 & +22:23:51.78 \\ 
1.2.20 & 11:49:35.752 & +22:23:51.00 \\
1.2.22 & 11:49:35.842 & +22:23:52.54 \\ 
1.2.23 & 11:49:35.709 & +22:23:52.45 \\ 
1.2.27 & 11:49:35.757 & +22:23:51.46 \\ 
1.3.1 & 11:49:36.820 & +22:24:08.77 \\ 
1.3.2 & 11:49:36.778 & +22:24:07.23 \\ 
1.3.3 & 11:49:36.906 & +22:24:07.37 \\ 
1.3.4 & 11:49:36.711 & +22:24:08.03 \\ 
1.3.5 & 11:49:36.921 & +22:24:09.24 \\ 
1.3.6 & 11:49:36.862 & +22:24:09.54 \\ 
1.3.7 & 11:49:36.809 & +22:24:10.34 \\ 
1.3.8 & 11:49:36.724 & +22:24:09.66 \\ 
1.3.9 & 11:49:36.888 & +22:24:08.68 \\ 
1.3.10 & 11:49:36.899 & +22:24:09.84 \\ 
1.3.11 & 11:49:36.944 & +22:24:08.69 \\ 
1.3.12 & 11:49:36.686 & +22:24:09.51 \\ 
1.3.13 & 11:49:36.850 & +22:24:09.14 \\ 
1.3.14 & 11:49:36.862 & +22:24:06.77 \\ 
1.3.15 & 11:49:36.784 & +22:24:09.33 \\ 
1.3.16 & 11:49:36.846 & +22:24:08.08 \\ 
1.3.17 & 11:49:36.807 & +22:24:08.25 \\ 
1.3.18 & 11:49:36.823 & +22:24:07.30 \\ 
1.3.19 & 11:49:36.789 & +22:24:10.21 \\ 
1.3.20 & 11:49:36.824 & +22:24:09.92 \\ 
1.3.21 & 11:49:36.730 & +22:24:08.84 \\ 
1.3.22 & 11:49:36.652 & +22:24:08.25 \\ 
1.3.23 & 11:49:36.749 & +22:24:10.17 \\ 
1.3.24 & 11:49:36.668 & +22:24:09.04 \\ 
1.3.25 & 11:49:36.911 & +22:24:08.30 \\ 
1.3.26 & 11:49:36.911 & +22:24:07.88 \\ 
1.3.27 & 11:49:36.796 & +22:24:09.81 \\ 
1.4.4 & 11:49:35.617 & +22:23:55.28 \\ 
1.4.7 & 11:49:35.542 & +22:23:53.69 \\ 
1.4.8 & 11:49:35.464 & +22:23:55.65 \\ 
1.4.8 & 11:49:35.681 & +22:23:53.62 \\ 
1.4.12 & 11:49:35.446 & +22:23:56.25 \\ 
1.4.19 & 11:49:35.501 & +22:23:54.33 \\ 
1.4.22 & 11:49:35.549 & +22:23:56.15 \\ 
1.4.23 & 11:49:35.437 & +22:23:55.20 \\ 
1.4.24 & 11:49:35.494 & +22:23:56.28 \\ 
1.4.27 & 11:49:35.563 & +22:23:54.19 \\ 
1.4.27 & 11:49:35.630 & +22:23:53.69 \\ 
1.4.28 & 11:49:35.597 & +22:23:54.32 \\ 
1.4.28 & 11:49:35.621 & +22:23:54.11 \\  
1.5.1 & 11:49:35.967 & +22:23:49.69 \\ 
1.5.2 & 11:49:36.031 & +22:23:49.93 \\ 
1.5.9 & 11:49:35.936 & +22:23:48.98 \\
1.5.16 & 11:49:35.999 & +22:23:49.11 \\ 
1.5.17 & 11:49:35.990 & +22:23:49.56 \\ 
1.5.18 & 11:49:36.044 & +22:23:49.35 \\ 
1.5.25 & 11:49:35.954 & +22:23:48.31 \\ 
1.5.26 & 11:49:35.991 & +22:23:48.17 \\ 
2.1.1 & 11:49:36.581 & +22:23:23.10 \\ 
2.2.1 & 11:49:37.450 & +22:23:32.92 \\ 
2.3.1 & 11:49:37.579 & +22:23:34.39 \\ 
3.1.1 & 11:49:33.772 & +22:23:59.36 \\ 
3.1.2 & 11:49:33.784 & +22:23:59.45 \\ 
3.1.3 & 11:49:33.825 & +22:23:59.50 \\ 
3.1.4 & 11:49:33.738 & +22:23:59.04 \\ 
3.1.5 & 11:49:33.795 & +22:23:59.67 \\ 
3.2.1 & 11:49:34.282 & +22:24:11.73 \\ 
3.2.2 & 11:49:34.252 & +22:24:11.10 \\ 
3.2.3 & 11:49:34.180 & +22:24:09.19 \\ 
3.2.4 & 11:49:34.326 & +22:24:12.73 \\ 
3.2.5 & 11:49:34.212 & +22:24:10.34 \\ 
3.3.1 & 11:49:36.279 & +22:24:25.88 \\ 
3.3.2 & 11:49:36.311 & +22:24:25.86 \\ 
3.3.3 & 11:49:36.394 & +22:24:25.74 \\ 
3.3.4 & 11:49:36.206 & +22:24:25.86 \\ 
3.3.5 & 11:49:36.339 & +22:24:25.92 \\ 
4.1.1 & 11:49:34.320 & +22:23:48.57 \\ 
4.2.1 & 11:49:34.651 & +22:24:02.65 \\ 
4.3.1 & 11:49:37.001 & +22:24:22.06 \\ 
5.1.1 & 11:49:35.940 & +22:23:35.02 \\ 
5.2.1 & 11:49:36.259 & +22:23:37.77 \\ 
6.1.1 & 11:49:35.930 & +22:23:33.16 \\ 
6.2.1 & 11:49:36.439 & +22:23:37.89 \\ 
7.1.1 & 11:49:35.750 & +22:23:28.82 \\ 
7.2.1 & 11:49:36.821 & +22:23:39.37 \\ 
7.3.1 & 11:49:37.819 & +22:24:04.47 \\ 
8.1.1 & 11:49:35.640 & +22:23:39.66 \\ 
8.2.1 & 11:49:35.950 & +22:23:42.16 \\ 
8.3.1 & 11:49:37.702 & +22:24:17.00 \\ 
9.1.1 & 11:49:36.890 & +22:23:52.03 \\ 
9.2.1 & 11:49:36.679 & +22:23:47.96 \\ 
9.3.1 & 11:49:36.010 & +22:23:37.89 \\ 
10.1.1 & 11:49:34.001 & +22:24:12.56 \\ 
10.2.1 & 11:49:33.799 & +22:24:09.53 \\ 
11.1.1 & 11:49:36.034 & +22:23:24.58 \\ 
11.2.1 & 11:49:36.965 & +22:23:34.42 \\ 
12.1.1 & 11:49:37.082 & +22:23:12.13 \\ 
12.2.1 & 11:49:37.920 & +22:23:20.61 \\ 
12.3.1 & 11:49:38.177 & +22:23:25.47 \\ 
13.1.1 & 11:49:38.484 & +22:23:19.50 \\ 
13.2.1 & 11:49:38.213 & +22:23:15.71 \\ 
13.3.1 & 11:49:37.495 & +22:23:07.33 \\ 
14.1.1 & 11:49:38.330 & +22:23:15.59 \\ 
14.2.1 & 11:49:38.371 & +22:23:16.21 \\ 
15.1.1 & 11:49:38.388 & +22:23:14.08 \\ 
15.2.1 & 11:49:38.695 & +22:23:18.46 \\ 
16.1.1 & 11:49:38.306 & +22:23:11.98 \\ 
16.2.1 & 11:49:38.899 & +22:23:20.60 \\  
\enddata
\tablenotetext{*}{ID1.ID2.ID3: ID1 - image system, ID2 - multiple image belonging to the image system ID1, ID3 - individual feature belonging to the multiple image ID2. }
\tablenotetext{*}{Images 1.X.Xs belong to the background spiral galaxy at $z = 1.4888$. Image 1.1.3 corresponds to the multiply-lensed SN Refdal. }
\end{deluxetable*}
\clearpage


\twocolumngrid
\bibliographystyle{aasjournal}
\bibliography{BH_references}


\end{document}